%% file: main.tex
\documentclass{aa}  

\usepackage{ulem} 
\usepackage{graphicx}
\usepackage{xcolor}
\usepackage{amsmath}	
\usepackage{amssymb}	
\usepackage{txfonts}
\usepackage{soul}
\usepackage[pdfauthor={Kristina Monsch, Giovanni Picogna, Barbara Ercolano, Wilhelm Kley},
pdftitle={Giant planet migration during the disc dispersal phase},
pdfsubject={XXX},
pdfkeywords={protoplanetary disks - planet-disk interactions - planets and satellites: gaseous planets - X-rays: stars – methods: numerical}]{hyperref} 

\newcommand\Msun{M_\odot}
\newcommand\Mjup{M_\mathrm{J}}
\newcommand\Mpl{M_{\rm p}}
\newcommand\Mdisc{M_\mathrm{d}}

\newcommand\Lx{L_\mathrm{x}}

\newcommand\yr{\mathrm{yr}}
\newcommand\Myr{\mathrm{Myr}}

\newcommand\ergs{\mathrm{erg\,s^{-1}}}
\newcommand\au{\mathrm{au}}

\newcommand\gcm{\mathrm{g\,cm^{-2}}}
\newcommand\keV{\mathrm{keV}}
\newcommand\xs{x_\mathrm{s}}
\newcommand\Mdotwind{\dot{M}_\mathrm{w}}
\newcommand\Sigmadotwind{\dot{\Sigma}_\mathrm{w}}

\newcommand\spock{\texttt{SPOCK}}
\newcommand\fargo{\texttt{FARGO}}

\begin{document} 

   \title{Giant planet migration during the disc dispersal phase}

   \author{Kristina Monsch \inst{1}
          \and
          Giovanni Picogna \inst{1}
          \and
          Barbara Ercolano \inst{1,2}
          \and
          Wilhelm Kley \inst{3}
          }

   \institute{Universit\"ats-Sternwarte, Ludwig-Maximilians-Universit\"at M\"unchen, Scheinerstr.~1, 81679 M\"unchen, Germany\\
        \email{[monsch, picogna, ercolano]@usm.lmu.de}
        \and
        Exzellenzcluster `Origins', Boltzmannstr.~2, 85748 Garching, Germany
   \and
      Institut f{\"u}r Astronomie und Astrophysik, Universit{\"a}t T{\"u}bingen, Auf der Morgenstelle 10, 72076 T{\"u}bingen, Germany\\
     \email{wilhelm.kley@uni-tuebingen.de}
   }

   \date{Received; accepted }

 
  \abstract
   {Transition discs are expected to be a natural outcome of the interplay between photoevaporation and giant planet formation. Massive planets reduce the inflow of material from the outer to the inner disc, therefore triggering an earlier onset of disc dispersal due to photoevaporation through a process known as Planet-Induced PhotoEvaporation (PIPE). In this case, a cavity is formed as material inside the planetary orbit is removed by photoevaporation, leaving only the outer disc to drive the migration of the giant planet.}
   {We investigate the impact of photoevaporation on giant planet migration and focus specifically on the case of transition discs with an evacuated cavity inside the planet location. This is important for determining under what circumstances photoevaporation is efficient at halting the migration of giant planets, thus affecting the final orbital distribution of a population of planets.}
   {For this purpose, we use 2D \fargo\ simulations to model the migration of giant planets in a range of primordial and transition discs subject to photoevaporation. The results are then compared to the standard prescriptions used to calculate the migration tracks of planets in 1D planet population synthesis models.}
   {The \fargo\ simulations show that once the disc inside the planet location is depleted of gas, planet migration ceases. This contradicts the results obtained by the impulse approximation, which predicts the accelerated inward migration of planets in discs that have been cleared inside the planetary orbit.}
   {These results suggest that the impulse approximation may not be suitable for planets embedded in transition discs. A better approximation that could be used in 1D models would involve halting planet migration once the material inside the planetary orbit is depleted of gas and the surface density at the 3:2 mean motion resonance location in the outer disc reaches a threshold value of $0.01\,\gcm$.}

   \keywords{protoplanetary disks -- planet-disk interactions -- planets and satellites: gaseous planets -- X-rays: stars -- methods: numerical}

   \maketitle
%
\input{01introduction.tex}
\input{02methods.tex}
\input{03results.tex}
\input{04discussion.tex}
\input{05conclusion.tex}

\begin{acknowledgements}
    We would like to thank Sijme-Jan Paardekooper for helpful comments, and the anonymous referee for a constructive report that improved the manuscript.
    We acknowledge the support by the DFG priority program SPP 1992 "Exploring the Diversity of Extrasolar Planets (DFG PR 569/13-1, ER 685/7-1) \& the DFG Research Unit ``Transition Disks'' (FOR 2634/1). 
    This research was supported by the Excellence Cluster ORIGINS, which is funded by the Deutsche Forschungsgemeinschaft (DFG, German Research Foundation) under Germany’s Excellence Strategy - EXC-2094 - 390783311. Part of the simulations have been carried out on the computing facilities of the Computational Center for Particle and Astrophysics (C2PAP).
\end{acknowledgements}

\bibliographystyle{aa} 
\bibliography{literature} 


\begin{appendix} 

\section{Details on the X-ray photoevaporation model}
\label{sec:appendix_PE}

\citet{Picogna+2019} present an updated X-ray photoevaporation model, which is based on a series of radiation-hydrodynamical simulations. 
They focus on modelling the photoevaporative winds of solar-type stars ($M_\star=0.7\,\Msun$) and show how these impact the dispersal of their surrounding disc at various stages of disc evolution. 

The photoevaporative surface mass-loss profile, $\Sigmadotwind(R)$, is described by

\begin{align}
\label{eq:Picogna_Sigmadotwind}
    \Sigmadotwind (R) =& \ln(10)\,\bigg( \frac{6a\ln(R)^5}{R\ln(10)^6} +\frac{5b\ln(R)^4}{R\ln(10)^5} +\frac{4c\ln(R)^3}{R\ln(10)^4}  \\ 
    \nonumber &+\frac{3d\ln(R)^2}{R\ln(10)^3}+\frac{2e\ln(R)}{R\ln(10)^2}
    +\frac{f}{R\ln(10)} \bigg)\, \frac{\Mdotwind(R)}{2\pi R}, 
\end{align} 
where $a = -0.5885$, $b = 4.3130$, $c= -12.1214$, $d = 16.3587$, $e = -11.4721$, $f = 5.7248$, $g = -2.8562$. The total mass-loss rate is derived via $\Mdotwind (R)=\int2\pi R\,\Sigmadotwind(R)\,\mathrm{d}R$ and yields:

\begin{align}
\label{eq:Picogna_MdotR}
    \Mdotwind(R) =& \,\Mdotwind(\Lx) \,\times\\ \nonumber 
    & \times 10^{a\log_{10}{R}^6 + b\log_{10}{R}^5 + c\log_{10}{R}^4 + d\log_{10}{R}^3 + e\log_{10}{R}^2 + f\log_{10}{R} + g},
\end{align}
where

\begin{equation}
\label{eq:Picogna_Mdotwind}
    \log_{10}(\Mdotwind(\Lx)/(M_\odot\, \mathrm{yr}^{-1})) = A_\textrm{L}  \exp{\left[\frac{(\ln{(\log_{10}(L_X))}-B_\textrm{L})^2}{C_\textrm{L}}\right]} + D_\textrm{L},
\end{equation}
and $A_\textrm{L} = -2.7326$, $B_\textrm{L} = 3.3307$, $C_\textrm{L} = -2.9868\cdot10^{-3}$ and $D_\textrm{L} = -7.2580$.
For transition discs with inner holes, Eq.~\ref{eq:Picogna_Sigmadotwind} becomes:

\begin{equation}
\label{eq:Picogna_Sigmadotwind_TD}
    \Sigmadotwind(R) = a b^{x} x^{c-1} (x \ln(b)+c)\ \frac{1.12\, \dot{M}(\Lx)}{2\pi R} ,
\end{equation}
where $a = 0.11843$, $b = 0.99695$, $c = 0.48835$, $x=R-R_\mathrm{gap}$, and $R_\mathrm{gap}$ describes the gap radius.

\begin{figure}
\centering
\includegraphics[width=\hsize]{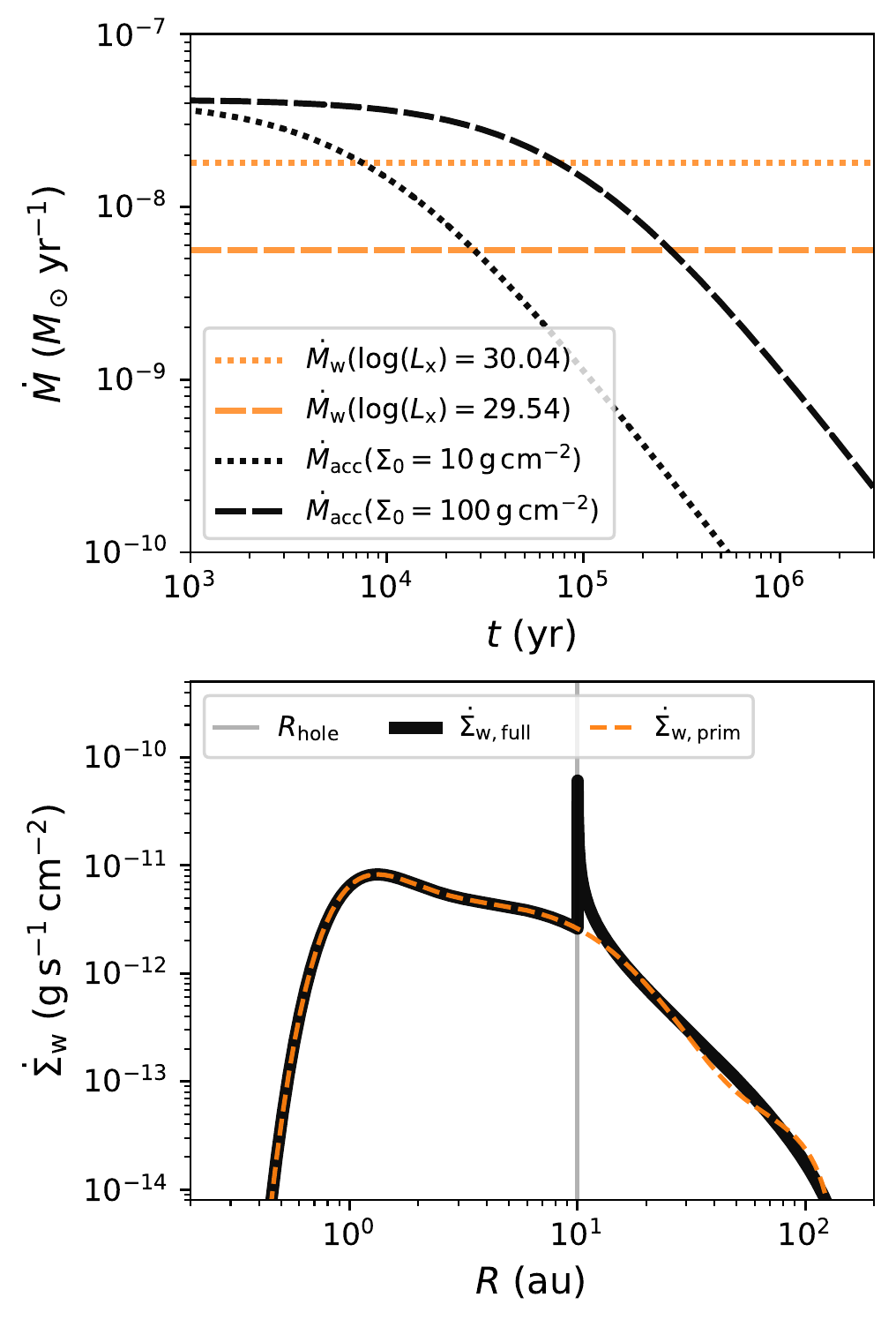}
  \caption{Comparison of the cumulative and surface mass-loss profile as predicted by the XPE model by \citet{Picogna+2019} to the purely viscous accretion rate onto the star. \textit{Top panel:} Viscous accretion rate at $1\,\au$ of the $\Sigma_0=10\,\gcm$ and $\Sigma_0=100\,\gcm$ discs as a function of time (black lines, cf. Eq.~35 in \citet{Hartmann+1998}) vs. the wind mass-loss rates due to photoevaporation for $\log_{10}(\Lx/\ergs)=29.54$ and $\log_{10}(\Lx/\ergs)=30.04$ (orange lines, cf. Eq.~\ref{eq:Picogna_Mdotwind}).
  \textit{Bottom panel:} Surface mass-loss profiles as a function of disc radius. The dashed line shows the full contribution of the primordial surface mass-loss profile, which is active before gap opening (Eq.~\ref{eq:Picogna_Sigmadotwind}). Assuming that photoevaporation has already opened a gap in the disc, with its outer radius $R_\mathrm{hole}$ lying at $10\,\au$, and the column density of the inner disc being less than $N_\mathrm{H}=2.5\times10^{22}\,\mathrm{cm}^{-2}$, the actual profile that is active is shown as the black solid line. }
     \label{fig:PE_profile}
\end{figure}

The top panel of Fig.~\ref{fig:PE_profile} compares the viscous accretion onto the star, $\dot{M}_\mathrm{acc} (R, t)$ \citep[see Eq.~35 in][]{Hartmann+1998}, evaluated at $1\,\au$ of the $\Sigma_0=10\,\gcm$ and $\Sigma_0=100\,\gcm$ discs with the wind mass-loss rate produced by photoevaporation for $\log_{10}(\Lx/\ergs)=29.54$ and $\log_{10}(\Lx/\ergs)=30.04$ (Eq.~\ref{eq:Picogna_Mdotwind}). 
It can be seen that in the early stages of disc evolution, the viscous accretion rates onto the star will exceed the photoevaporative mass-loss rates. Depending on the initial disc mass and the X-ray luminosity of the star, the accretion rates will drop below the wind mass-loss rate at a given time, so that photoevaporation can open an annular gap and start clearing the disc from the inside out. 
In the bottom panel of Fig.~\ref{fig:PE_profile}, the surface mass-loss profile as a function of disc radius is shown. 
The dashed line shows the primordial profile (Eq.~\ref{eq:Picogna_Sigmadotwind}), which is active before photoevaporation has opened a gap. As soon as gap opening has taken place and the column density inside this gap has decreased sufficiently, the stellar X-rays can directly irradiate the outer disc, so that the photoevaporation profile switches to the transition disc one outside of the gap (Eq.~\ref{eq:Picogna_Sigmadotwind_TD}). The total surface mass-loss profile is therefore a combination of the primordial and the transition disc one that will intersect at the gap radius, $R_\mathrm{hole}$. 

\begin{figure}
\centering
\includegraphics[width=\hsize]{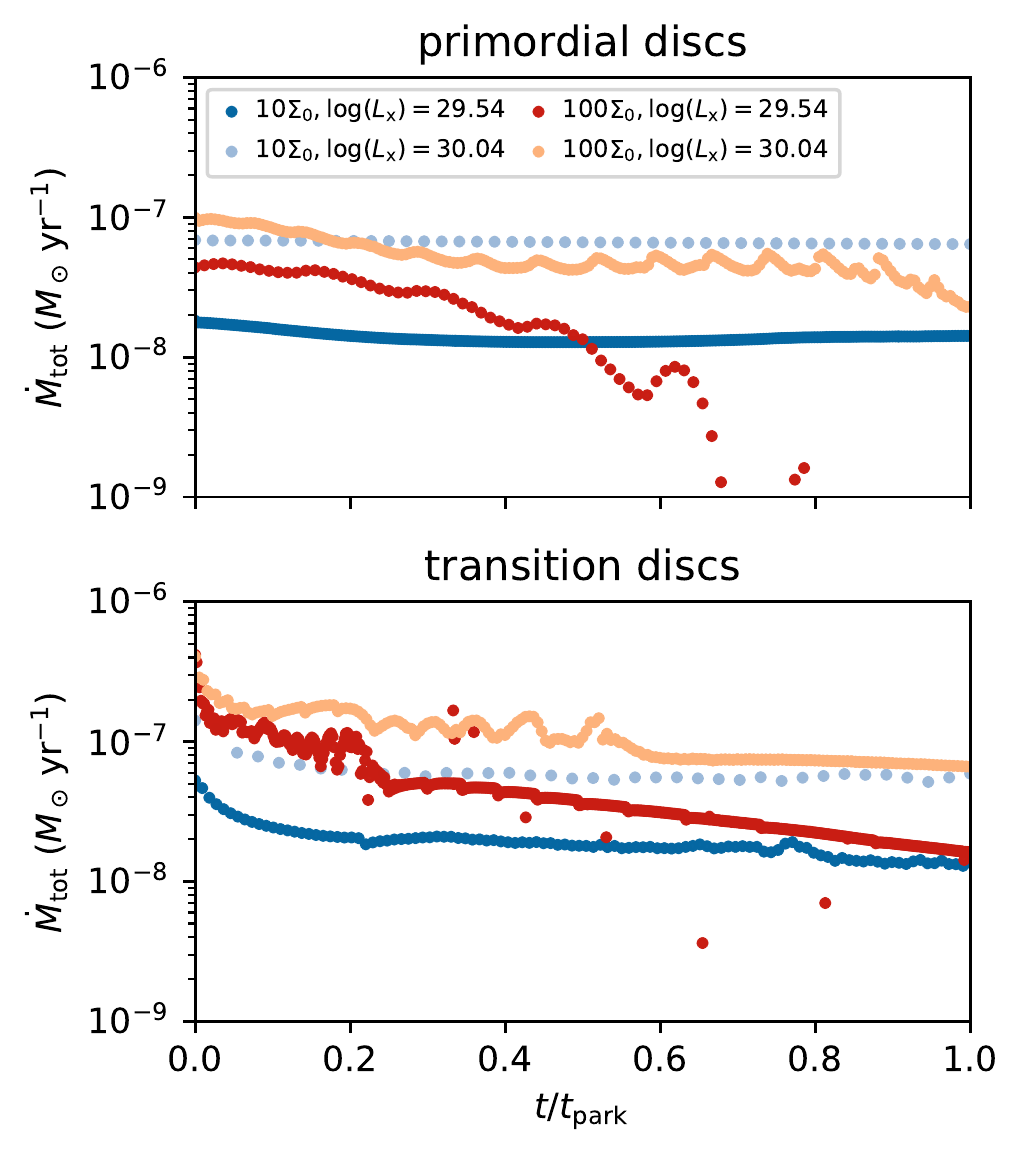}
  \caption{Total mass-loss rate, $\dot{M}(t)$, of the discs modelled with \fargo\ as a function of $t/t_\mathrm{park}$, where $t_\mathrm{park}$ is the time it takes the planet to get parked in each simulation. The lines are plotted starting from the times at which the planet is released and photoevaporation is activated. The total mass-loss rates are averaged over 100 orbits. The top panel shows $\dot{M}_\mathrm{tot}$ in the primordial discs, while the bottom one corresponds to transition discs.  }
     \label{fig:mass_loss}
\end{figure}

The total mass-loss rate (i.e. including viscous accretion and photoevaporation) of the discs modelled with \fargo\ are shown in Fig.~\ref{fig:mass_loss}. As expected, higher mass-loss rates are reached both in the low- and high-mass discs with higher X-ray luminosities of the host star due to the more vigorous winds that are more efficient in removing disc material. This trend is independent of the underlying disc structure, however, in the case of transition discs, even higher absolute values of the mass-loss rate compared to the primordial discs are present. Rather than being absorbed by the inner disc, the X-rays can directly irradiate the outer disc in these cases, so that the stronger, direct photoevaporation profile (Eq.~\ref{eq:Picogna_Sigmadotwind_TD}) becomes immediately active. This results in higher total mass-loss rates and the faster removal of the outer disc compared to the primordial discs.

Overall, the total mass-loss rates stay relatively constant over the disc lifetime, especially for the low-mass discs. For the high-mass discs with $\Sigma_0=100\,\gcm$ the viscous accretion rates are higher (see the discussion in Sect.~\ref{sec:results_fulldiscs}) and consequently a stronger decrease in $\dot{M}_\mathrm{tot}$ with time is observed. 
This is especially prominent for $\log_{10}(\Lx/\ergs)=29.54$, where $\dot{M}_\mathrm{tot}$ quickly reaches values below $10^{-9}\,\Msun\,\mathrm{yr}^{-1}$. Due to the high disc mass and low X-ray luminosity, photoevaporation is not strong enough to overcome viscous accretion before the planet migrates all the way to the host star (see Fig.~\ref{fig:comparison_SPOCK_FARGO}). The mass-loss rate therefore follows a trend similar to that of the viscous accretion rate observed in Fig.~\ref{fig:PE_profile} (black, dashed line) as the viscous accretion rate will exceed the photoevaporation mass-loss rate for a longer time. 
For the transition discs mostly the wind mass-loss rate is contributing to the total mass-loss of the disc. The evacuated inner cavity formed by PIPE prevents any accretion onto the star, and therefore all gas that crosses the planetary orbit is immediately photoevaporated away.

\section{The role of viscosity}
\label{sec:appendix_alpha}

\begin{figure}
\centering
\includegraphics[width=\hsize]{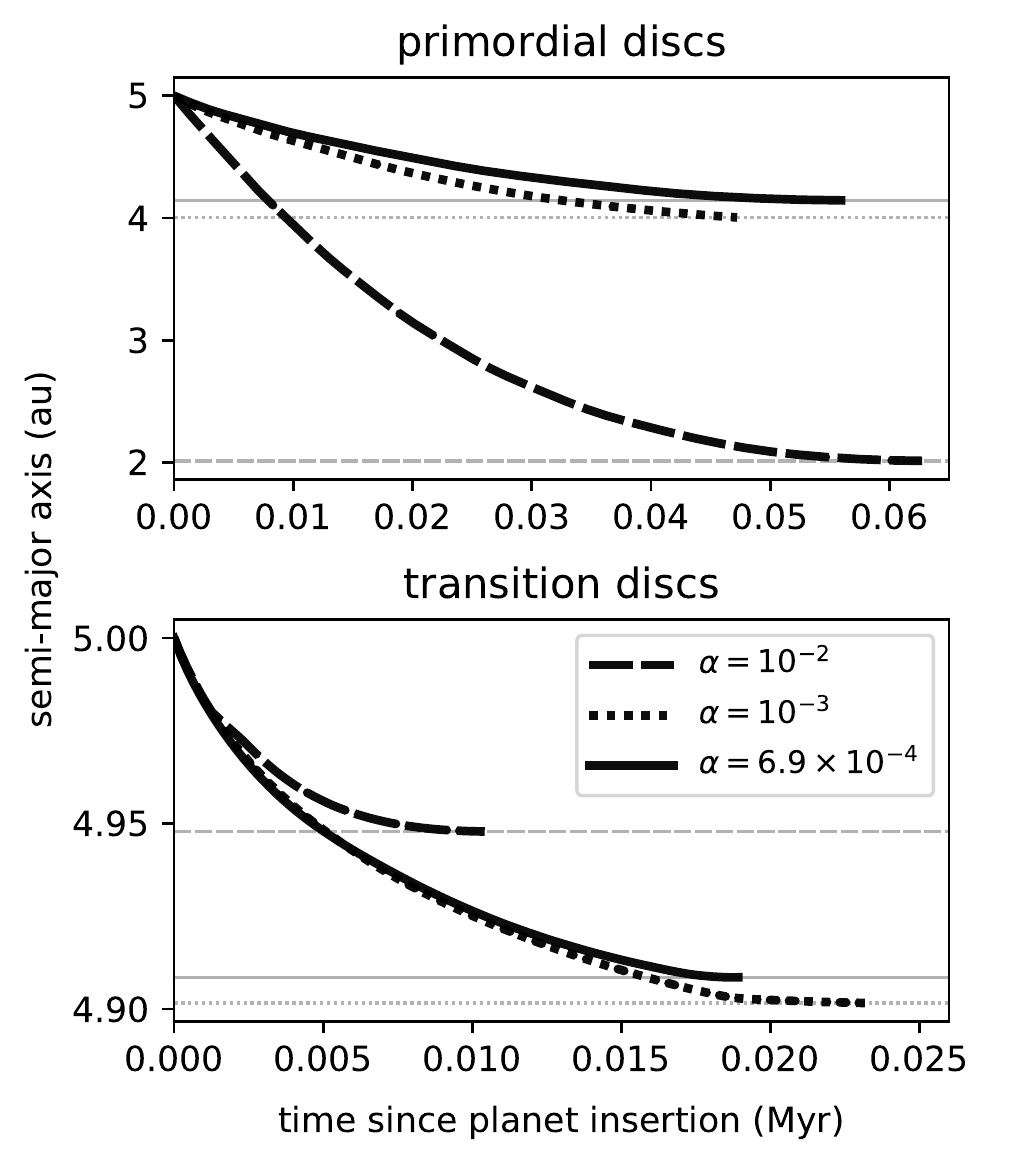}
  \caption{Semi-major axis evolution of planets embedded in primordial discs (top panel) vs. planets embedded in transition discs (bottom panel), modelled with \fargo\ for different $\alpha$-parameters. }
     \label{fig:comp_alpha}
\end{figure}

It was discussed in Sect.~\ref{sec:results_fulldiscs} that photoevaporation is more efficient in parking planets in low-mass discs. In comparison, the planets embedded in the higher-mass discs with $\Sigma_0=100\,\gcm$ were effectively not parked due to planet-disc interactions or photoevaporation, but reached the inner boundary of the radial domain. 
The magnitude of planet migration in photoevaporating discs however crucially depends on the viscosity applied in these models. 
Fig.~\ref{fig:comp_alpha} shows the semi-major axis evolution of a planet embedded in the $\Sigma_0=10\,\gcm$ and $\log_{10}(\Lx/\ergs)=29.54$ primordial and transition discs for three different $\alpha$-viscosities, namely $\alpha=[10^{-2}, 10^{-3}, 6.9\times10^{-4}]$. 

If the viscosity is lower, the migration speed of the planet will become slower, as can be observed in the full disc simulations. The reason for this is that the gaps that are carved by the planets, are wider and deeper in low-viscosity discs, therefore the torques acting on the planet will be reduced. In contrast, for the higher viscosity cases, the planetary gap is partially refilled by inflowing gas, producing stronger torques and leading to the faster inward migration of the planet.
For the transition discs opposite behaviour is observed, that is that the planet embedded in the disc with $\alpha=10^{-2}$ is parked the earliest. However, the difference in the final parking locations compared to the other modelled disc viscosities is only marginal.
The reason for this is that with higher disc viscosity, material will accumulate on a faster timescale close to the planet location in the outer disc. As was shown in Fig.~\ref{eq:impulse_torques}, one-sided torques near the gap edge will create strongly positive torques, effectively preventing any further inward migration of the planet. 
Consequently, with higher disc viscosity more material will be present at the gap edge, resulting in the formation of even stronger positive torques that lead to an earlier stopping of planet migration compared to transition discs with lower viscosity.
In conclusion, while the choice of $\alpha$ may affect the final parking location of the planet, it does not change the conclusion that planet migration should cease once the disc reaches the transitional phase, in which the disc inside the planet is depleted of gas.

\section{Torque fluctuations}
\label{sec:appendix_bumps}

\begin{figure*}[ht!]
\centering
\includegraphics[width=\hsize]{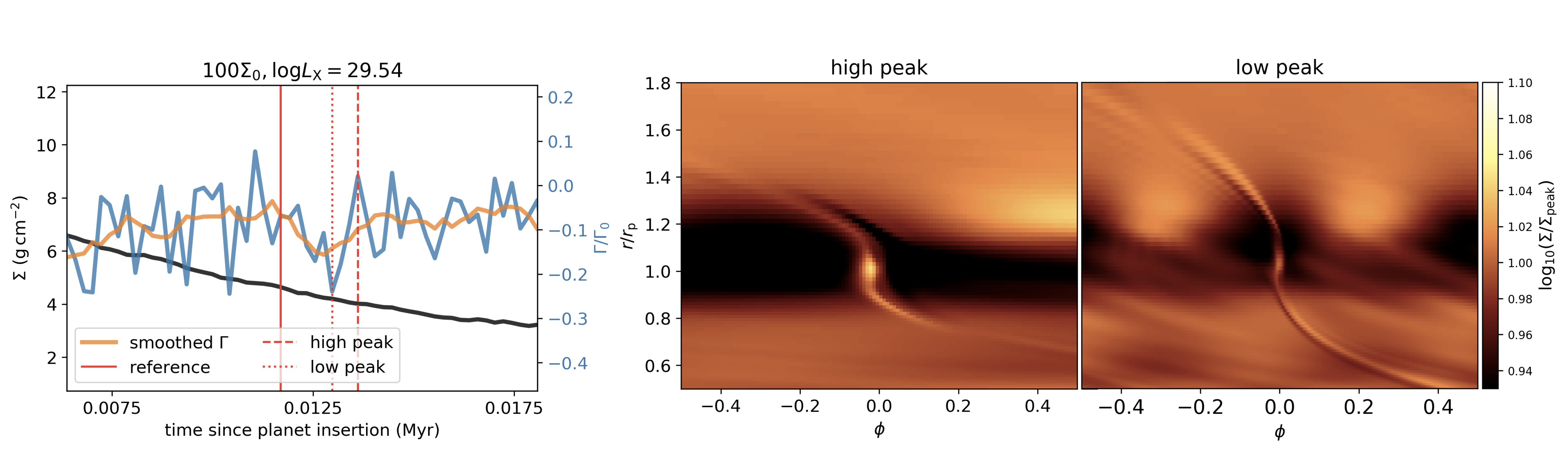}
  \caption{Asymmeteries in the gas surface density close to the planet as a cause for the torque oscillations observed in Fig.~\ref{fig:torques}. \textit{Left panel}: Torque evolution as a function of time for the low X-ray luminosity case. The blue line corresponds to the ``raw'' torque directly obtained from \fargo, while the orange line shows the smoothed torque using the previously described Savitzky-Golay filter. The black line corresponds to the azimuthally averaged surface density profile at the planet location. The torques were normalised by the normalisation factor $\Gamma_0$ (cf. Eq.~\ref{eq:gamma0}).
  \textit{Right panel}: Corresponding 2D surface density distribution for the high and low torque peaks with respect to the reference step.}
     \label{fig:torque_bumps}
\end{figure*}

The torque evolution discussed in Sect.~\ref{sec:results_torques} showed strong oscillations with time, especially for the high-mass discs ($\Sigma_0 = 100\,\gcm$) for which disc-planet interactions were more important. The case with low X-ray luminosity ($\log_{10}(\Lx/\ergs)=29.54$) showed in particular short-term oscillations while for the high X-ray case the variation was much longer with time.
In Figure~\ref{fig:torque_bumps} we show the surface density distribution close to the planet location at two time steps where the torque had a low or high peak with respect to a reference step, for the low X-ray luminosity case. The surface density shows strong interaction with vortices developing in the outer gap edge that are well in agreement with the short-term oscillations. The long-term variation observed in the high X-ray flux case can be instead related to the strong interaction between the stellar irradiation and the gap edge outside the planet location, which affects the torque acting onto the planet.

\end{appendix}
\end{document}

%% file: 01introduction.tex
\section{Introduction}
\label{sec:introduction}

Giant planet migration is regarded as a natural outcome of the gravitational interaction between a forming planet and the surrounding gas in the planet-forming disc. While the existence of this process was predicted more than 40 years ago \citep{LinPapaloizou1979, GoldreichTremaine1979, GoldreichTremaine1980}, its importance for planet formation theories was only realised with the first discoveries of the so-called hot Jupiters \citep[e.g.][]{MayorQueloz1995}. These Jupiter-like planets with orbital periods of $P< 15\,\mathrm{d}$ \citep{Wang+2015} are considered to be direct evidence of planet migration having taken place. Their in situ formation at such small distances to their host stars is unlikely \citep[see, however,][and \citet{Boley+2016} for alternative explanations]{Batygin+2016}, suggesting that these planets generally formed in the outer parts of the planet-forming disc and migrated inwards during their evolution \citep[e.g.][]{DawsonJohnson2018}.

In the last few years, scattered light and (sub-)millimetre surveys have provided observations of circumstellar discs with different substructures, such as rings \citep[e.g.][]{ALMA_HLTau+2015, vanBoekel+2017}, spirals \citep[e.g.][]{Benisty+2015, Dong+2018, Muro-Arena+2020}, or shadows \citep[e.g.][]{Benisty+2017, Walsh+2017}. 
While there is a range of mechanisms that can potentially explain these substructures, such as gas pressure bumps at ice lines \citep[e.g.][]{Zhang+2015, Okuzumi+2016, Owen2020}, photoevaporation \citep[e.g.][]{Ercolano+2017}, or non-ideal magnetohydrodynamic (MHD) effects \citep[e.g.][]{Hu+2019}, the deep gaps and other substructures in many of these systems are commonly believed to have been carved by nascent planets \citep[e.g.][]{PaardekooperMellema2004, PicognaKley2015, DipierroLaibe2017, Zhang+2018, Veronesi+2020}. 
The discovery of the two young, accreting protoplanets PDS\,70b and PDS\,70c \citep{Keppler+2018, Mueller+2018, Haffert+2019} proved that giant planets can indeed form early and that gaps and rings in so-called transition discs may offer potential indirect evidence for embedded planets \citep{Rosotti+2016, Sanchis+2020} and possibly also for their migration \citep{Meru+2019, Nazari+2019, Perez+2019}.

Independently of when exactly giant planet formation sets in, the dispersal of the gas-phase of planet-forming discs sets not only a strict upper limit to the planet formation timescale \citep{Pollack+1996}, but also to planet migration as this is the result of the exchange of angular momentum between the planet and the gas in the disc.
As the migration timescale of giant planets \citep[typically $\sim10^5\,\yr$, cf.][]{KleyNelson2012} has been shown to be much shorter than the observed lifetimes of planet-forming discs \citep[typically $\sim 10^6~\yr$, cf.][]{Haisch+2001, Mamajek+2009, Fedele+2010, Ribas+2014, Ribas+2015}, this elucidates the necessity of formulating planet formation theories in combination with disc dispersal mechanisms.

It is commonly believed that planet-forming discs disperse via a combination of viscous accretion and disc winds, possibly launched due to internal photoevaporation by the host star. Theoretical models predict that viscous accretion dominates for most of the disc lifetime, until the accretion rates can no longer match the mass-loss rate due to disc winds, which will finally disperse the disc quickly on a timescale of a few hundred thousand years from the inside out \citep{Alexander2014_PP6, ErcolanoPascucci2017}. The evolution of young stars on colour-colour diagrams of nearby star-forming regions is indeed consistent with the prediction of a fast inside-out dispersal phase \citep{Ercolano+2009b, Ercolano+2011, Koepferl+2013}. 

Models show that the gas is heated and finally unbound, predominantly by soft X-rays ($0.1\,\keV \leq E \leq 1\,\keV$) emitted by the host star so that over a range of disc radii a vigorous disc wind is established and eventually a gap opens up in the disc. This decouples the inner from the outer disc, so that the former drains viscously onto the central star, while the latter is photoevaporated away from the inside out \citep{Ercolano+2008, Ercolano+2009a, Owen+2010, Owen+2011, Owen+2012}. The influence of extreme ultraviolet radiation (EUV, $13.6\,\mathrm{eV}\leq E \leq 0.1\,\mathrm{keV}$) on the final mass-loss rates is expected to be negligible due to the fact that EUV photons are readily absorbed by atomic hydrogen in the disc atmosphere, and thus have a much smaller penetration depth \citep[][but see also \citet{WangGoodman2017} and \citet{Nakatani+2018}]{Ercolano+2008}. The notion that the EUV luminosities impinging on protoplanetary discs are low compared to soft X-rays has also been observationally confirmed \citep{Pascucci+2014}. 
Internal photoevaporation rates driven by far ultraviolet emission (FUV, $6\,\mathrm{eV} \leq E \leq 13.6\,\mathrm{eV}$) yield comparable mass-loss rates of $\sim 10
^{-8}\,\Msun\,\mathrm{yr}^{-1}$ to the XEUV-models \citep{Gorti+2009, GortiHollenbach2009}, depending on the assumption made for the small grain population in the disc atmospheres. However, no hydrodynamical calculations exist to date for internal photoevaporation driven by FUV, and thus the mass-loss profiles, $\dot{\Sigma}_{\mathrm{wind}}$, and mass-loss rates, $\dot{M}_{\mathrm{wind}}$, still need to be confirmed by future calculations.

It has been numerically shown that disc dispersal via photoevaporative winds can strongly affect the migration of giant planets and therefore ultimately leave an imprint in their final semi-major axis distribution \citep{AA09, AP12, ER15, Jennings+2018}. These works also show that the final orbital distribution of giant planets strongly depends on the radial profile of the mass-loss rates. It is thus in principle possible, within the limitation of the numerical models \citep[see the discussions in][and \citet{Jennings+2018}]{ER15}, to observe an imprint of the disc dispersal phase in the present-day giant planet orbital distribution. 
\citet{Monsch+2019} made a first attempt at finding possible signatures of X-ray driven photoevaporation (XPE) in the observational exoplanet data, presumably established in the disc dispersal phase and driven by the highly energetic radiation emitted by the host star. 
To this aim, they looked for a possible link between the X-ray properties of stars hosting giant planets and their corresponding semi-major axis distribution.
They assembled a catalogue of the stellar X-ray luminosities and found a suggestive void in the $\Lx$ versus semi-major axis plane. This void could be qualitatively explained as a consequence of disc dispersal via XPE, which stops giant planet migration at a given place in the disc for a given range of $\Lx$. However, due to the small sample size, the statistical significance of this feature could not be proven with the data set at that time.
\citet{Monsch+2019} argued that, without having a significant increase in observational data points, accurate numerical models are required to predict the location and/or size of this gap a priori, in order to prove its statistical significance. 

For such purposes, 1D planet population synthesis models are an ideal tool as they allow us to assess the impact of a range of different initial conditions on the final architecture of planetary systems \citep[cf.][and \citet{Mordasini2018}, for reviews]{Benz+2014}. However, several simplifications have to be assumed in such models to be able to run a statistically significant amount of simulations, ultimately limiting their predictive power. In particular, approximating a three-dimensional problem to 1D is the largest source of error since it requires the description of geometrically complex problems like angular momentum exchange between the gas disc and the planet in one dimension. This is generally achieved through prescriptions derived from more complex multi-dimensional hydrodynamical calculations.

One of these prescriptions is the so-called impulse approximation \citep{LinPapaloizou1979}, which is used to approximate the torques that are exerted by the gaseous disc on a planet. Since it yields the correct scaling as more complicated treatments \citep[e.g.][]{Trilling+1998}, it has been widely used in 1D planet population synthesis approaches to model the migration of planets embedded in gaseous discs \citep[cf.][and \citet{KleyNelson2012}, for reviews]{LubowIda2010}.
In our case, this regime corresponds to so-called type~II migration, in which the planet is entrained in the viscous flow of the gas, therefore migrating at the viscous accretion speed.
However, the classical paradigm of type~II migration has recently been questioned, suggesting that the migration rate of the planet is not locked to the disc's viscous evolution \citep[][and \citet{Kanagawa+2018}; however, see also \citet{Robert+2018}]{Duffell+2014, DuermannKley2015}.
While the impulse approximation yields reasonable results for many disc-planet systems, its accuracy for more complex configurations, in which the 2D structure of the disc may be of relevance, still needs to be tested and adapted as, for example, by \citet{LiuOrmelLin2017}, who studied the migration process of Super Earths trapped at the magnetospheric cavity of the planet-forming disc and are therefore subject to one-sided torques.

Using 2D hydrodynamical \fargo\ simulations, \citet{Rosotti+2013, Rosotti+2015} studied the interplay of giant planet formation and disc dispersal via XPE, concluding that giant planets may trigger the faster onset of disc dispersal by reducing the mass inflow into the inner disc. This effect was referred to as Planet-Induced PhotoEvaporation (PIPE). The disc around TW~Hya may be a possible example of PIPE taking place as the planet, which might have carved the outer rings in the disc observed with ALMA \citep{Andrews+2016}, may have also triggered the onset of photoevaporation that has cleared the disc at small radii \citep{Ercolano+2017}. However, \citet{Rosotti+2013, Rosotti+2015} did not investigate in detail which effect this would have on the migration process of planets. 

In this paper, we extend the work by \citet{Rosotti+2013} and investigate how the interplay of XPE and planet formation may affect giant planet migration. The aim is to assess whether the commonly employed impulse approximation in 1D models can correctly describe the migration of giant planets in photoevaporating discs. 
For this purpose, we perform 2D \fargo\ simulations to compare migration tracks of giant planets embedded in primordial discs to those embedded in transition discs for which the disc inside the planet has been cleared (for example by PIPE). These are then compared to results obtained from 1D simulations with the same input parameters that employ the impulse approximation.

The paper is structured as follows. In Section~\ref{sec:methods}, we describe the 1D and 2D simulations used for this study. Section~\ref{sec:results} summarises and discusses the results, while Section~\ref{sec:discussion} highlights the limitations of our numerical model. In Section~\ref{sec:conclusion} we draw our conclusions and propose how planet migration in transition discs formed by PIPE could be handled in 1D planet population synthesis approaches.

%% file: 02methods.tex
\section{Methods}
\label{sec:methods}

To investigate the impact of photoevaporation on giant planet migration, we made use of both 1D and 2D simulations to compare migration tracks from the impulse approximation to those obtained from the full treatment of the underlying disc torques in two dimensions. Our study therefore only concerns the late phases of disc evolution, meaning when disc dispersal is taking place, assuming that the planet has already formed. 
In each of these approaches, we considered both primordial and transition discs formed by PIPE.
Considering the former, we sought to study the impact of photoevaporation on the overall migration history of the planet, while considering the latter, we specifically investigated the process of giant planet migration in discs with an evacuated cavity inside the planetary orbit, so that the planet is only subject to one-sided torques acting from a rather massive outer disc.

\subsection{X-ray photoevaporation model}
\label{sec:methods_PE}

We employed the XPE profile with an EUV component from the recent radiation-hydrodynamical calculations by \citet{Picogna+2019} and refer the reader to that work for more details on the underlying physical model. It improves on previous work by \citet{Owen+2010, Owen+2011, Owen+2012} by parameterising the temperature as a function of local gas properties and the column density to the star. This results in an almost twice higher mass-loss rate compared to \citet{Owen+2012} and differences in the radial profile of the mass-loss rate. \citet{Woelfer+2019} present X-ray photoevaporation models for carbon-depleted discs, predicting even more vigorous winds with radially extended profiles. Studying the migration of giant planets in X-ray photoevaporated, carbon-depleted discs is, however, beyond the scope of this paper and will be attempted in future work. 
While photoevaporative winds are expected to be dusty \citep{Franz+2020}, the remaining dust in the disc is not likely to affect giant planet migration significantly \citep[see][for a study on the impact of dust on the migration of low-mass planets]{Benitez-LlambayPessah2018}.

The strength of the photoevaporative mass-loss rate is primarily determined by the stellar X-ray luminosity, which typically reaches highly elevated levels for young T~Tauri stars compared to their main-sequence counterparts \citep[e.g.][]{FeigelsonMontmerle1999}. We used the X-ray luminosity function determined for the Taurus molecular cloud by \citet{Guedel+2007} and scale this to a stellar mass of $0.7\,\Msun$, with a resulting median of $\Lx=1.1\times10^{30}\,\ergs$.
To cover the intrinsic scatter of X-ray luminosities of T~Tauri stars for a given stellar mass, we considered two values for the X-ray luminosity in the simulations presented here, namely $\Lx=2.7\times10^{29}\,\ergs$ and $\Lx=1.1\times10^{30}\,\ergs$.

\citet{Picogna+2019} present different mass-loss profiles for primordial and transition discs, which we included in our calculations. 
These profiles are described in detail in Appendix~\ref{sec:appendix_PE}. 
For most of the disc lifetime, the viscous accretion rate onto the star ($\sim 10^{-8}\,\Msun\,\yr^{-1}$ at $1\,\Myr$, cf. \citet{Hartmann+1998}) exceed the mass-loss rate due to photoevaporation, which for the above stated X-ray luminosities of $\Lx=2.7\times10^{29}\,\ergs$ and $\Lx=1.1\times10^{30}\,\ergs$, reaches values of $5.6\times10^{-9}\,\Msun\,\yr^{-1}$ and $1.8\times10^{-8}\,\Msun\,\yr^{-1}$. 
After the viscous accretion rate drops below the wind mass-loss rate, photoevaporation will open a gap in the disc, and cut off the inner disc from further mass supply from the outer disc. 
While the inner disc drains viscously onto the central star, the outer disc is irradiated directly by the stellar X-rays as soon as the column density of the remaining material inside the gap becomes less than the maximum X-ray penetration depth of roughly $2.5\times10^{22}\,\mathrm{cm}^{-2}$ \citep{Ercolano+2009a, Picogna+2019}. At this point the switch between the primordial and transition disc profile is performed in our models.

\subsection{1D model using impulse approximation}
\label{sec:methods_spock}

To model the migration of a giant planet in an XEUV-irradiated disc following the impulse approximation, we used the 1D viscous evolution code \spock\ \citep{ER15}. We followed a similar setup as described by \citet{ER15} and refer the reader to their work for more details on the numerical model, which we will only briefly summarise. 

We modelled a planet-forming disc evolving under the influence of viscosity and X-ray photoevaporation from the host star with a mass of $0.7\,\Msun$ and 
assumed an initial disc surface density profile of 

\begin{equation}
\label{eq:sigma_spock}
    \Sigma(R, t=0) = \frac{\Mdisc(t=0)}{2\pi R_\mathrm{s} R}\,\exp\left( -\frac{R}{R_\mathrm{s}} \right),
\end{equation}
where $\Sigma(R,t)$ is the gas surface density of the disc and $R$ the distance from the star.
Eq.~\ref{eq:sigma_spock} follows from the self-similarity solution of the diffusion equation using a time-independent power-law scaling of the disc radius with the kinematic viscosity, $\nu\propto R^\gamma$, assuming $\gamma=1$ \citep[cf.][]{LyndenBellPringle1974, Hartmann+1998}.
We used an initial disc mass of $\Mdisc(0)=0.07\,\Msun$ with a disc scaling radius of $R_\mathrm{s}=18\,\au$, which defines the initial disc size and sets the viscous timescale of the disc \citep{AA09}.
Following the standard $\alpha$-prescription \citep{ShakuraSunyaev1973}, viscosity is defined as $\nu=\alpha c_\mathrm{s}H$, where $c_\mathrm{s}$ is the sound speed of the gas, $H$ the disc scale height and $\alpha$ a dimensionless parameter.
We used $\alpha=6.9\times10^{-4}$, which is consistent with models treating realistic hydrodynamical turbulence \citep{Picogna+2018}, resulting in a viscous timescale of $t_\nu=7\times10^{5}\,\yr$ at $R_\mathrm{s}$. This combination of viscosity and disc scaling radius were chosen such that a population of discs subject to XPE from stars with an observationally motivated X-ray luminosity (cf. Sect.~\ref{sec:methods_PE}) have a mean disc lifetime of a few Myr as suggested by observations of young stellar clusters \citep{Haisch+2001, Mamajek+2009, Fedele+2010, Ribas+2014, Ribas+2015}. 

The evolution of the planet-disc system can be described following the equation:

\begin{equation} 
\frac{\partial \Sigma}{\partial t} = \frac{1}{R}\frac{\partial}{\partial R}\left[ 3R^{1/2} \frac{\partial}{\partial R}\left(\nu \Sigma R^{1/2}\right) - \frac{2 \Lambda \Sigma R^{3/2}}{(G M_\star)^{1/2}}\right] - \dot{\Sigma}_{\mathrm {w}}(R,t),
\label{eq:sigma_evol}
\end{equation}
where the first term on the right hand side describes the viscous evolution of the disc \citep{LyndenBellPringle1974}, the second term deals with the migration of the planet due to the torques exerted from the gas in the disc \citep[e.g.][]{LinPapaloizou1986}, and finally, the last term is the mass-loss due to photoevaporation (described by Eq.~\ref{eq:Picogna_Sigmadotwind} and Eq.~\ref{eq:Picogna_Sigmadotwind_TD}). Here, $M_\star$ is the stellar mass, $G$ the gravitational constant and $\Lambda$ is the rate of angular momentum transfer per unit mass from the planet to the disc.
Eq.~\ref{eq:sigma_evol} is discretised on a grid of $n_\mathrm{r}=1000$ radial cells equispaced in $R^{1/2}$, extending from $0.04\,\au$ to $10^4\,\au$. 

A fully formed planet of $\Mpl=1\,\Mjup$ was then inserted at $5\,\au$ into the disc once it had reached a surface density of $\Sigma_0=10\,\gcm$ or $100\,\gcm$ at the planet insertion location. 
At this point the resolution was increased to $n_\mathrm{r}=4000$ and our simulations were divided into two sets - primordial and transition discs. In the former, the planet was inserted into the full disc, and was left to evolve undisturbed following disc-planet interactions. In the latter, as soon as the planet was inserted, we set the surface density to the floor value of $\Sigma_\mathrm{floor}=10^{-8}\,\gcm$ inside the planet location to manually create transition discs with an evacuated inner cavity, such as would be formed by PIPE.

Planetary accretion was modelled following Eq.~5 in \citet{VerasArmitage2004} and took place in both sets of our simulations, regardless of the inner disc being cleared or not. 
The planet then migrated at a rate given by the impulse approximation \citep{LinPapaloizou1979, LinPapaloizou1986}:

\begin{equation}
\frac{\mathrm{d}a}{\mathrm{d}t} = -\left(\frac{a}{G M_\star} \right )^{1/2} \left(\frac{4\pi}{M_\mathrm{p}}\right) \int_{R_\mathrm{in}}^{R_\mathrm{out}}{R\Lambda\Sigma}\mathrm{d}R. 
\end{equation}
The angular momentum input $\Lambda$ is poorly known, and we therefore followed the formalism introduced by \citet{Armitage+2002}, which is a slight modification to the original one by \citet{LinPapaloizou1986}:

\begin{equation}
\Lambda(R,a) = \left\{ \begin{array}{ll}
- \frac{q^2 G M_\star}{2R} \left(\frac{R}{\Delta_{\mathrm p}}\right)^4 & \textrm{if } \, R < a\\
\frac{q^2 G M_\star}{2R} \left(\frac{a}{\Delta_{\mathrm p}}\right)^4 & \textrm{if } \,R > a.\\
\end{array}\right. 
\label{eq:impulse_torques}
\end{equation}
Here, $q=\Mpl/M_\star$ is the mass ratio between the planet and the star, $a$ is the semi-major axis of the planetary orbit (assumed to be circular) and $\Delta_p$ is given by

\begin{equation}
 \Delta_\mathrm{p} = \max (H, |R-a| ),
\end{equation}
where $H$ is the disc scale height. $\Delta_\mathrm{p}$ corresponds to the impact parameter that ensures that material closer than one disc scale height is excluded from the torque calculation.  

The simulations were stopped once the disc had been dispersed, or the planets reached $a=0.15\,\au$ as we do not attempt to model the interaction with the magnetospheric cavity.
Table~\ref{table:1} summarises the initial conditions used for the 1D model, which are the same for both the primordial and transition disc simulations. The only parameters that were varied in the 1D simulations are the initial surface density of the disc, $\Sigma_0$, and the X-ray luminosity of the star, $\Lx$. 

\begin{table}
\caption{Initial conditions for the \spock\ simulations described in Sect.~\ref{sec:methods_spock}.}             
\label{table:1}      
\centering                          
\begin{tabular}{l l}        
\hline\hline                 
\spock\ parameter & value\\
\hline                        
$M_\star~(\Msun)$ & $0.7$ \\
$\Mdisc~(\Msun)$ & $0.07$ \\
$\alpha$ & $6.9\times 10^{-4}$ \\
$R_\mathrm{s}$ (au) & $18.$ \\
$H/R$ & 0.1 \\
$\Lx~(\ergs)$ & $2.7\times10^{29}$, $1.1\times10^{30}$ \\
$\Mpl~(\Mjup)$ & $1.$ \\
$a_0$~(au) & $5.$ \\
$\Sigma_0$ ($\gcm$) & 10, 100 \\
$\Sigma_\mathrm{floor}$ ($\gcm$) & $10^{-8}$ \\
$R_\mathrm{in}$ (au) & 0.04 \\ 
$R_\mathrm{out}$ (au) & $10^4$ \\  
\hline                                   
\end{tabular}
\end{table}

\subsection{2D \fargo\ simulations}
\label{sec:methods_fargo}

\begin{figure}
\centering
\includegraphics[width=\hsize]{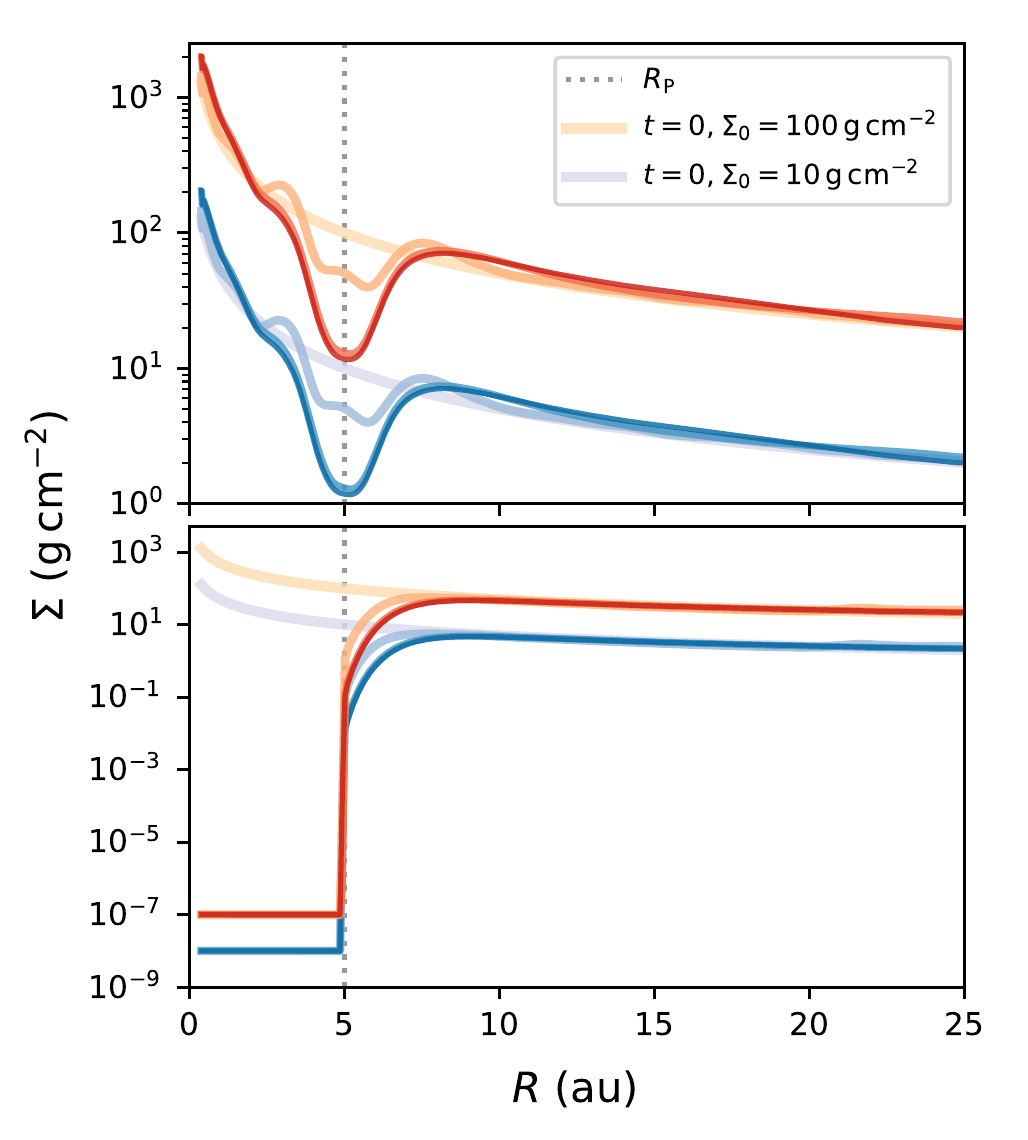}
  \caption{Azimuthally averaged gas surface density evolution of the primordial (top) and transition discs (bottom) modelled with \fargo. The different lines correspond to snapshots at orbit 0, 100, 1000, and 1500 for the primordial disc at orbit 0, 10, 150, and 200 for the transition disc. During this time, the planet position was kept fixed and photoevaporation was switched off.}
     \label{fig:sigma_comp}
\end{figure}

To study the interaction of a giant planet embedded in an X-ray irradiated disc in 2D, we used a modified version of the hydrodynamical grid code \texttt{FARGO} \citep{Masset2000}, in which we included XPE.
Before the planet is inserted, no detailed calculations of the disc structure are required due to the axis-symmetry of the system, which substantially reduces the computational expense of the simulations. However, once the planet is formed, this axis-symmetry will break as the planet induces spiral density waves and eventually carves a gap. Therefore, following the approach described in \citet{Rosotti+2013}, we used the surface density profile of the discs in the 1D simulations at the time of planet insertion as an input for the \fargo\ simulations.

We adopted a cylindrical polar coordinate system $(r, \phi, z)$ centred on the star. 
To implement a similar disc structure into \fargo\ as in the 1D simulations, we made use of the following relation (which is the default one in \fargo):

\begin{equation}
\label{eq:sigma_fargo}
    \Sigma(r) = \Sigma_0 \left ( \frac{R}{l_0} \right )^{-p}, 
\end{equation}
where $\Sigma_0$ is the initial surface density at the base length unit of $l_0=5\,\au$ and $p=1$ is the slope of the surface density profile. We note that this profile is different than the one described by Eq.~\ref{eq:sigma_spock} that was used in the 1D simulations. However, we tested that Eq.~\ref{eq:sigma_fargo} correctly reproduces the surface density profile from \spock\ in the smaller radial domain modelled with \fargo\ at the time of planet insertion. 
Following the 1D simulations, we considered two different values of $\Sigma_0$, namely $10\,\gcm$ and $100\,\gcm$, while we kept the remaining parameters, unless otherwise stated, fixed to their default values. Further we assumed locally isothermal discs with a constant aspect ratio of $h=0.1$ and a flaring index of $f=0.25$, resulting in a mildly flared disc: $H/R=0.1 R^{0.25}$.

We modelled the disc from $0.4\,\au$ to $25\,\au$ with a resolution of $n_r=256$ cells in radial and $n_\theta=388$ cells in azimuthal direction using logarithmic spacing, which yielded approximately square cells at the planet location. This resolution was chosen in order to allow for a long integration time, while being able to probe the parameter space. We further performed a higher resolution simulation for comparison that gave similar results.
We applied `viscous outflow' boundary conditions at both boundaries of the radial grid to impose a steady-state accretion flow both from the outer boundary into the disc and from the inner disc onto the central star \citep[for details, see Eq.~11 in][]{Kley+2008}. The magnitude of the flow's radial velocity at the corresponding boundary radius is determined by the viscosity parameter $\alpha$ that was set to the same value of $\alpha=6.9\times10^{-4}$ as in the 1D runs. 

Photoevaporation was included as a sink-term in the continuity equation \citep{MoeckelArmitage2012, Rosotti+2013}:

\begin{equation} 
    \frac{\partial \Sigma}{\partial t} + \nabla \cdot (\Sigma \mathbf{v}) = - \dot{\Sigma}_\mathrm{w} (R, t).
\end{equation}
The mass was removed from the disc surface density at the beginning of each time step and to prevent negative surface densities from arising, we used a floor value of $10^{-9}\times\Sigma_0$ throughout the disc.
We considered the same X-ray luminosities as in the 1D model, namely $\Lx=2.7\times10^{29}\,\ergs$ and $\Lx=1.1\times10^{30}\,\ergs$ for a stellar mass of $0.7\,\Msun$. However, to extract the impact of photoevaporation onto planet migration, we also performed control simulations without photoevaporation. 
Equivalently to the approach described in Sect.~\ref{sec:methods_spock}, we generated transition discs in \fargo\ by setting the surface density inside the planet location to the above stated floor value at the beginning of each transition disc simulation.

\begin{table}
\caption{Initial conditions that were used for the 2D \fargo\ simulations described in Sect.~\ref{sec:methods_fargo}. The parameters are given in the name convention employed in the \fargo\ executable files.}
\label{table:2}      
\centering                          
\begin{tabular}{l l }        %
\hline\hline                 
\fargo\ parameter & value \\    
\hline                        
\texttt{Sigma0} $(\gcm)$ & 10, 100\\
\texttt{SigmaSlope} & 1. \\
\texttt{SigmaFloor} ($\Sigma_0$) & $10^{-9}$ \\
\texttt{AlphaViscosity} & $6.9\times10^{-4}$\\
\texttt{AspectRatio} & 0.1 \\
\texttt{FlaringIndex} & $0.25$ \\
\texttt{l0}~(au) & 5. \\
\texttt{m0}~($\Msun$) & 0.7 \\
\texttt{mu} & 2.35\\
\texttt{Nrad} & 256 \\
\texttt{Nsec} & 388 \\
\texttt{Rmin} ($l_0$) & 0.08 \\
\texttt{Rmax} ($l_0$) & 5. \\
\texttt{RadialSpacing} & Logarithmic \\
\texttt{InnerBoundary} & Viscous \\
\texttt{OuterBoundary} & Viscous \\
\texttt{Adiabatic} & No \\
\texttt{PlanetMass} ($\Mjup$) & 1. \\
\texttt{PlanetDistance} ($\au$) & 5.\\

\hline                                   
\end{tabular}
\tablefoot{If a parameter is not specified, the default value is used. We note that \fargo\ employs code units, which are set by the base mass $m_0$, the base length $l_0$ and the time per orbit $t_0=\sqrt{l_0^3/(Gm_0)} = 1/\Omega_0$, where $G$ is the gravitational constant and $\Omega_0$ the Keplerian orbital frequency.}
\end{table}

The planet was inserted at $5\,\au$ into the disc and gradually grew to its final mass of $1\,\Mjup$ following a mass-taper function $\Mpl(t) = m_\mathrm{taper}\times1\,\Mjup$, where
 
\begin{equation} 
    m_\mathrm{taper} = \sin^2 \left( \frac{t}{4t_\mathrm{ramp-up}} \right),
\end{equation}
and $t_\mathrm{ramp-up} = 10$ orbits is the ramp-up time. We note that after this step, the planet was not allowed to accrete any more disc material. The implications of this approach on the reliability of our results are discussed in detail in Sect.~\ref{sec:discussion_leakage}.
The planet was kept fixed for 1500 orbits in the primordial discs and 200 orbits in the transition discs to 1) allow the planet to carve a deep gap in the surface density and 2) to stabilise the torques acting on the planet before it is allowed to migrate and photoevaporation is switched on. This is illustrated in Fig.~\ref{fig:sigma_comp}, which shows the azimuthally averaged surface density evolution of the primordial and transition discs at different snapshots during the first orbits, in which the planet was kept fixed. It can be seen that the disc, especially towards the outer boundaries, had reached a stable state long before the planet was released.
Table~\ref{table:2} summarises the initial conditions used for the \fargo\ simulations. 

%% file: 03results.tex
\section{Results and discussion}
\label{sec:results}

\subsection{Migration in primordial discs subject to XPE}
\label{sec:results_fulldiscs}

\begin{figure*}[t!]
\centering
\includegraphics[width=\hsize]{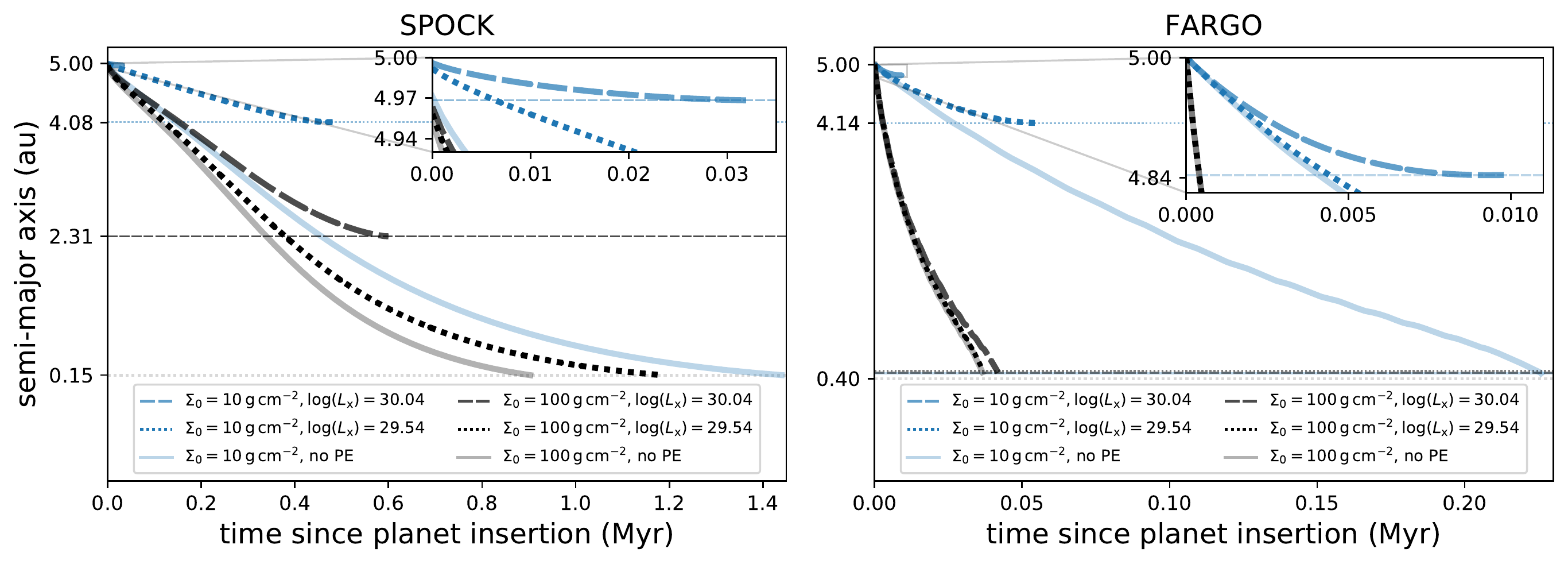}
  \caption{Comparison of the semi-major axis evolution of a $1\,\Mjup$ planet in the primordial discs for different X-ray luminosities: $\log (\Lx/\ergs)=30.04$ (dashed), $\log (\Lx/\ergs)=29.54$ (dotted), and vs. no photoevaporation (solid), computed with \spock\ (left panel) and \fargo\ (right panel). The blue lines correspond to an initial surface density of $\Sigma_0=10\,\gcm$ and the black lines to $\Sigma_0=100\,\gcm$. The horizontal lines are drawn at the final planet parking locations. For \spock, the simulations are stopped once the planets reach $0.15\,\au$, while for \fargo\ the inner grid boundary lies at $0.4\,\au$. The upper right panels zoom into the parameter space to show the evolution of the $\log (\Lx/\ergs)=30.04$ simulations as their total disc lifetimes are significantly shorter than for the other simulations.}
     \label{fig:Lx_fullcomp}
\end{figure*}

Fig.~\ref{fig:Lx_fullcomp} shows the semi-major axis evolution of the planets embedded in primordial discs, computed from \spock\ using the impulse approximation (left) and from the full 2D treatment using \fargo\ (right). Each line represents a different setup sampled from the two different initial disc masses of $\Sigma_0=10\,\gcm$ (`low-mass disc') and $\Sigma_0=100\,\gcm$ (`high-mass disc') as well as X-ray luminosities of $\log_{10}(\Lx/\ergs)=29.54$ (`low $\Lx$') and $\log_{10}(\Lx/\ergs)=30.04$ (`high $\Lx$'), while the remaining parameters were kept constant.

Due to the vigorous photoevaporative winds, planets embedded in the low-mass discs with high X-ray luminosities are effectively immediately parked once they are inserted into the disc. This can be seen in the subplots in the upper right corners of Fig.~\ref{fig:Lx_fullcomp}, that zoom in the corresponding parameter space. For this setup, the final parking locations of the planets as well as the timescales are comparable both in the 1D and 2D approach. In \spock, the planet migrates by about $0.03\,\au$ within $3\times10^{4}\,\yr$, while in \fargo\ it takes the planet only $10^4\,\yr$ to get parked at $4.84\,\au$. 
If no photoevaporation-driven mass-loss is applied, all planets migrate up to the inner radial boundary of the corresponding radial grid for both modelled disc masses, demonstrating the efficiency of X-ray photoevaporation at stopping giant planet migration. 

The most significant differences in the final planet parking locations for both approaches can be observed for $\log_{10}(\Lx/\ergs)=29.54$.
In \spock, the planet is parked at $4.08\,\au$ after approximately $0.5\,\Myr$ for the low-mass disc, but for $\Sigma_0=100\,\gcm$ it migrates up to the inner radial boundary within $1.2\,\Myr$. This is explained by the fact that for the latter approach, the accretion rate through the disc exceeds the mass-loss rate due to photoevaporation, which is then unable to remove the material around the planet that is responsible for the torques that cause its inward migration. 
Qualitatively similar behaviour can be observed in the \fargo\ simulations, in which the planet is stopped at $4.14\,\au$ after only $0.06\,\Myr$ in the low-mass disc, while for the higher-mass disc significantly faster migration compared to \spock\ is observed. However, also in this case the planet migrates up to the inner radial boundary in less than $0.05\,\Myr$.

Consequently, while the simulations in \spock\ and \fargo\ both qualitatively confirm that planets are expected to migrate less with higher X-ray luminosity of the host star and that higher migration rates are obtained for more massive discs
\citep[as the torques acting on the planet directly scale with the disc's surface density, cf.][]{LinPapaloizou1986}, they also show that the absolute timescales differ strongly in the two approaches.
While in \fargo\ the planets are parked in less than $0.3\,\Myr$, in \spock\ they span a broader range between a few $10^5\,\yr$ to $\sim1.4\,\Myr$.
These differences can be mainly related to the different extent and surface density profile of the disc, but more importantly, to the different disc evolution in both approaches. While it is possible to model the entire disc-planet system for its full lifetime in 1D, in \fargo\ one has to limit the simulations to a smaller region close to the planet due to the higher computational expense of the 2D simulations.
The simplifications and uncertainties involved in the 1D approach will be addressed in more detail in Sect.~\ref{sec:discussion}, however we additionally refer the reader to the discussion in \citet{Rosotti+2013}.
Fig.~\ref{fig:Lx_fullcomp} therefore shows that a direct comparison between the timescales in 1D and 2D simulations should be treated with caution. 

As mentioned above, photoevaporation is clearly more effective in the low-mass discs. Here, the migration history of the planets and the final parking location are most dramatically affected by this process. 
This finding is in agreement with recent results by \citet{WiseDodsonRobinson2018}, who showed that the impact of photoevaporation on parking planets is negligible, if the ratio of the planet to disc mass is small, so that $M_\mathrm{disc} \gg \Mpl$. 
This is obvious as in high-mass discs the accretion rates exceed the photoevaporation rates such that any material removed by photoevaporation is readily replaced so that the net effect on the torques is then negligible. In this context it is however important to address the role of viscosity. The accretion rate onto the star is not directly driven by the disc mass itself, but by the disc viscosity, which in turn is set by the $\alpha$ parameter, the disc scale height, $H$, and the sound speed of the gas, $c_\mathrm{s}$. In our simulations, $\alpha$ was kept constant while we explored a given range of disc masses. Thus, the mass accretion rate scales with the surface density of the disc, which in our simple case can be directly related to the disc mass. Therefore, for a given value of $\alpha$, the higher-mass discs accrete at higher rates than the lower-mass discs and lie well above the photoevaporative mass-loss rate. This means that the latter has negligible influence on the surface density evolution of the high-mass disc, and thus on the torques exerted on the planet. 
For completeness, we present the effect of different $\alpha$-viscosities on the migration of the planets for the case example of $\Sigma_0=10\,\gcm$ and $\log_{10} (\Lx/\ergs)=29.54$ in Appendix~\ref{sec:appendix_alpha}. We find that applying different viscosities in our simulations does not change the overall conclusion of this paper. However, it will certainly change the migration rates that are observed for the individual planets and impact their final parking location.

\begin{figure}
\centering
\includegraphics[width=\hsize]{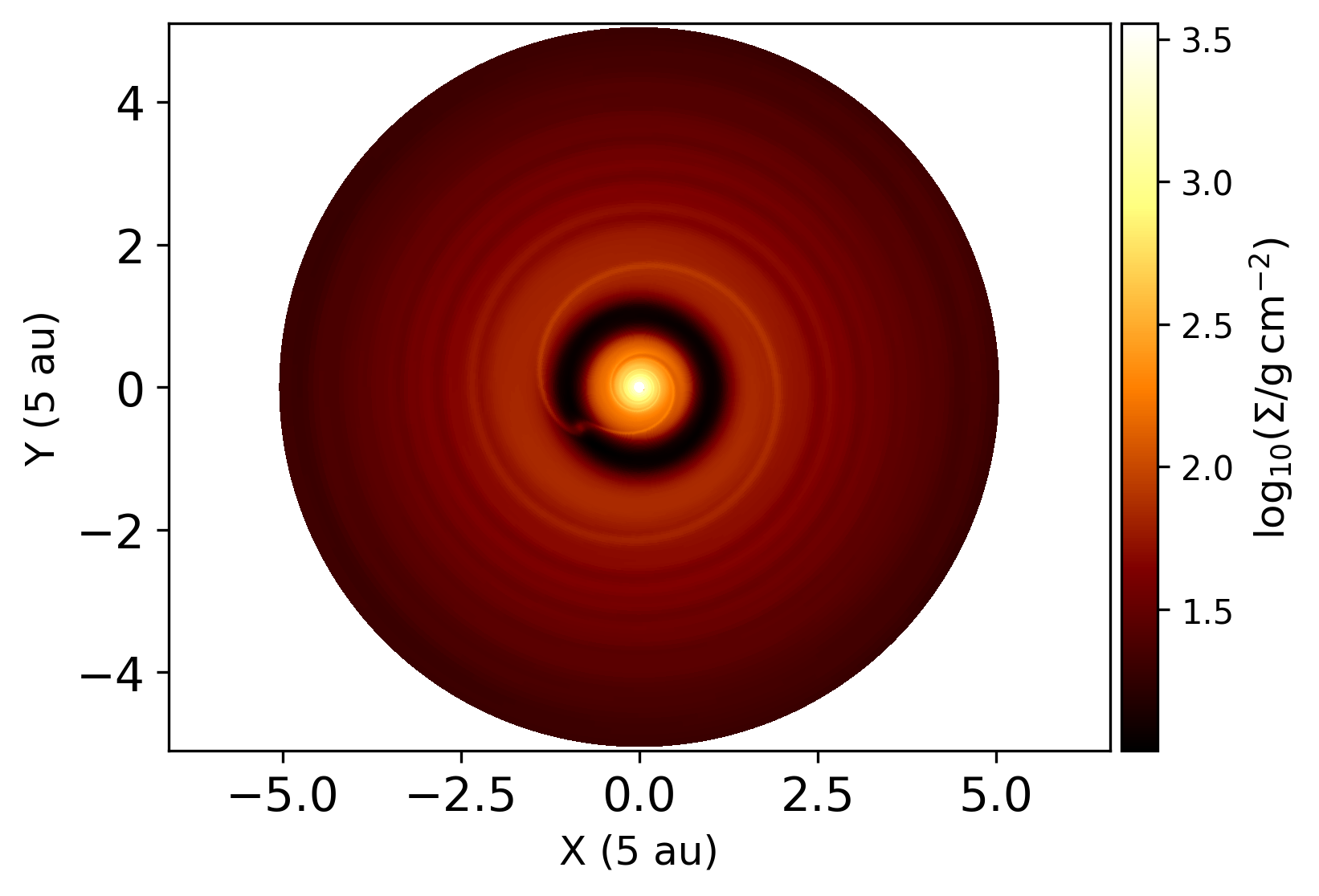}
  \caption{2D surface density distribution of the $\Sigma_0=100\,\gcm$ primordial disc, determined from \fargo. The surface density is plotted at orbit 1500, at which the planet is released and photoevaporation is activated.}
     \label{fig:2D_density}
\end{figure}

In conclusion, for primordial discs the impulse approximation employed in \spock\ gives a reasonably good match for the final parking locations in the low-mass discs compared to the more realistic treatment with \fargo. There is, however, a larger discrepancy for the higher-mass discs, where even in the control simulations without photoevaporation, the migration is faster in the 2D calculations. 
The main reason for this discrepancy is the different disc evolution between the two approaches as well as the interaction between the planet and the spiral arms that develop in the 2D simulations.  
These provide local enhancements to the torques acting on the planet that cannot be reproduced and accounted for in axisymmetric 1D calculations. The spiral arms induced by the planet can be seen in Fig.~\ref{fig:2D_density}, which shows the 2D gas surface density distribution of the primordial disc with $\Sigma_0=100\,\gcm$ at orbit 1500, that is right before the planet is released and photoevaporation is activated.

\subsection{Migration of planets in primordial versus transition discs}
\label{sec:results_fullvsonesided}

\begin{figure*}
\centering
\includegraphics[width=\hsize]{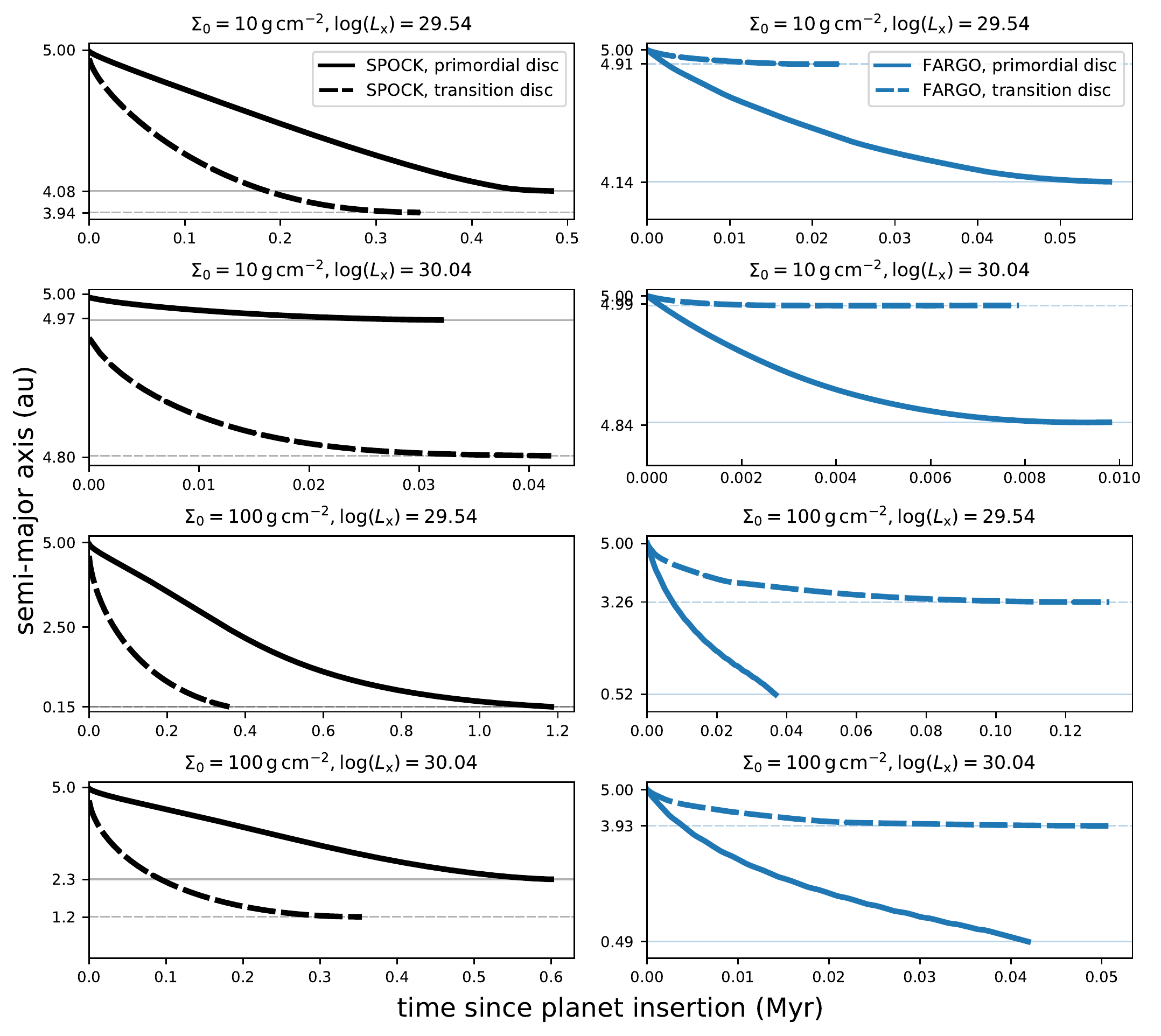}
  \caption{Comparison of the semi-major axis evolution and the final parking locations for different initial disc masses and X-ray luminosities, computed from \spock\ (left panels) and \fargo\ (right panels). The solid lines show the evolution for the planets embedded in primordial discs, while the dashed lines correspond to planets embedded in transition discs. The horizontal lines are drawn at the corresponding planet parking location from each simulation.}
     \label{fig:comparison_SPOCK_FARGO}
\end{figure*}

Fig.~\ref{fig:comparison_SPOCK_FARGO} shows a comparison of the semi-major axis evolution for the primordial and transition discs, computed with \spock\ and \fargo. 
Using the impulse approximation implemented in \spock, planets embedded in transition discs generally migrate farther inside than they do in primordial discs, except for the case of $100\,\gcm$ and $\log_{10}(\Lx/\ergs)=29.54$, where for both setups the planet is stopped at the inner radial grid. 
However, in all of the scenarios modelled in 1D, the planets embedded in transition discs show accelerated inward migration compared to the planets in primordial discs.
Naively, this behaviour may be expected due to the missing counteracting effect of the inner disc, so that the relatively massive outer disc can push the planet inside at an increased migration speed.
However, it becomes immediately apparent from the more realistic \fargo\ simulations that the migration of planets subject to one-sided torques only exerted from an outer disc is not treated correctly using the impulse approximation in 1D.
The \fargo\ results show that the planet migrates the farthest inside, if it is embedded in a primordial disc. 
In contrast to that, the planet migrates only weakly if it is embedded in a transition disc. The magnitude of migration depends however on the initial disc mass and the strength of the photoevaporative winds.
While for $\Sigma_0=10\,\gcm$ only insignificant migration can be observed for both modelled X-ray luminosities, for the higher-mass discs, the planet migrates up to $3.26\,\au$ for $\log_{10}(\Lx/\ergs)=29.54$ and up to $3.93\,\au$ for $\log_{10}(\Lx/\ergs)=30.04$. 
The \fargo\ simulations therefore show that planets in transition discs generally migrate significantly less than they do in primordial discs. 

In summary, when comparing the migration tracks between \spock\ and \fargo\ for the transition discs only, it becomes apparent that in the latter, planets generally migrate significantly less than they do in \spock, while for the primordial discs no strong deviation is found, except for $100\,\gcm$ and $\log_{10}(\Lx/\ergs)=30.04$, where the slower migration in \spock\ allows photoevaporation to deplete the disc more strongly. 
As a consequence, planet migration is slowed down even more and the planet is parked before it can reach the inner boundary.

\subsection{Torques acting on the planet}
\label{sec:results_torques}

\begin{figure*}
\centering
\includegraphics[width=\hsize]{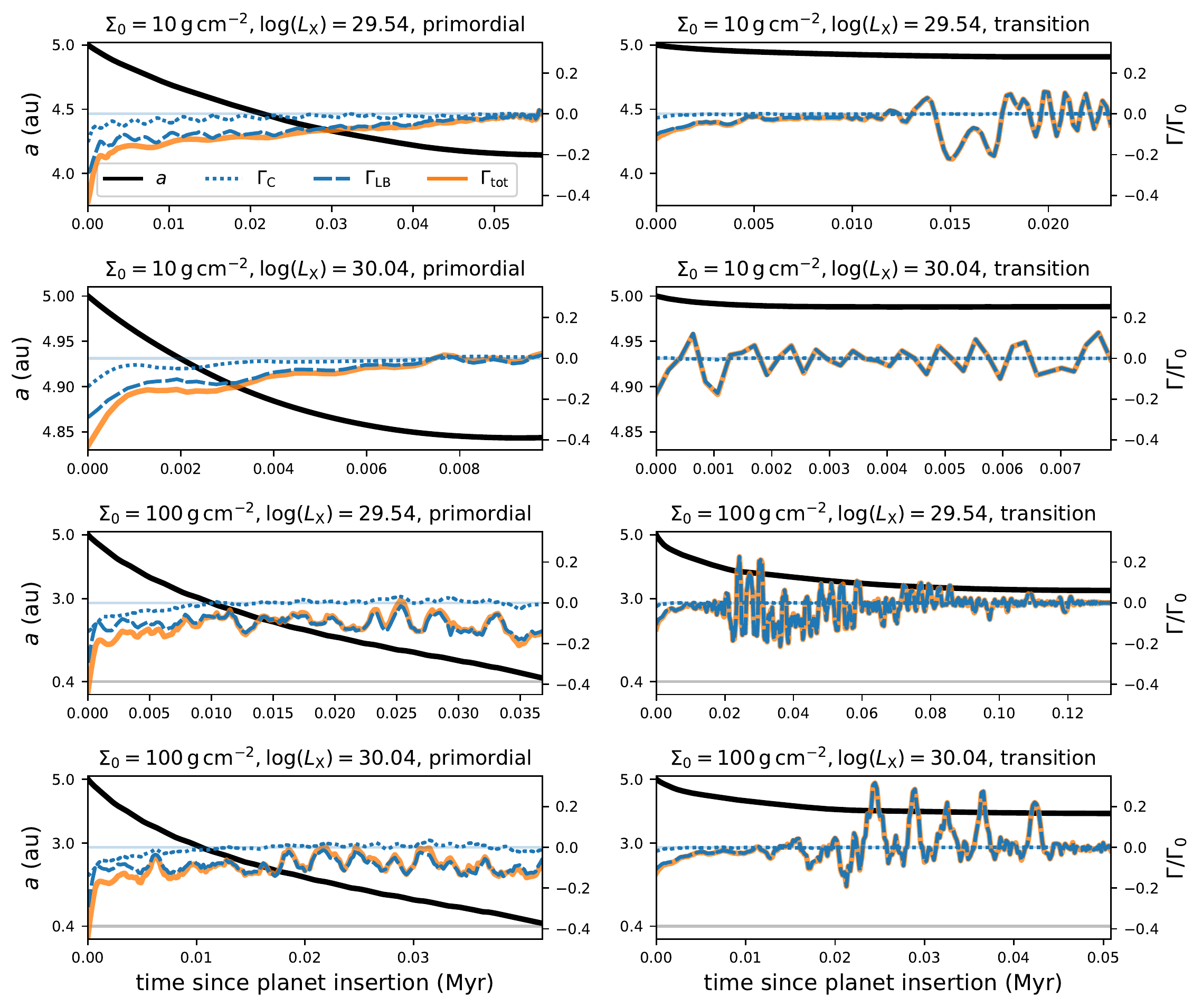}
  \caption{Comparison of the semi-major axis evolution (black solid) of the planet with respect to the corotation ($\Gamma_\mathrm{C}$, blue dotted), Lindblad ($\Gamma_\mathrm{LB}$, blue dashed), and total torques ($\Gamma_\mathrm{tot} = \Gamma_\mathrm{C}+ \Gamma_\mathrm{LB}$, orange solid) in the primordial (left panels) and transition discs (right panels), calculated from \fargo.
  The torques were divided by $\Gamma_0$, which is the unperturbed torque at the planet location (cf. Eq.~\ref{eq:gamma0}).}
     \label{fig:torques}
\end{figure*}

To understand the origin of the different migration rates of planets embedded in primordial and transition discs,  Fig.~\ref{fig:torques} shows the evolution of the total torques ($\Gamma_\mathrm{tot}$) acting on the planet, together with the individual contribution of the corotation ($\Gamma_\mathrm{C}$) and Lindblad ($\Gamma_\mathrm{LB}$) torques as a function of time. These are directly compared to the semi-major axis evolution of the planet.

The corotation torques are determined within the co-orbital region $\xs$ \citep[also known as horseshoe half-width, cf.,][]{PaardekooperPapaloizou2009}, spanning a region close to the planetary orbit from $a-\xs \leq a \leq a+\xs$. 
The width of $\xs$ is given by

\begin{equation}
    \xs=1.68a\sqrt{\frac{q}{h}},
\end{equation}
where $a$ corresponds to the planet semi-major axis, $q=\Mpl/M_\star$ is the planet to stellar mass ratio and $h$ the disc aspect ratio \citep{PaardekooperPapaloizou2009}.
The contribution of the Lindblad torques are then calculated from the remaining parts of the disc. 
The total torque is therefore the sum of the individual corotation and Lindblad torques and the net difference between the torques interior and exterior to the planet determine its final migration rate.
For most planet-disc configurations, the net difference between the torques becomes negative so that it drives inward migration \citep{Ward1997}, which is also the case in our simulations where no significant outward migration was observed.

The cumulative torques as a function of time were normalised by 

\begin{equation}
\label{eq:gamma0}
    \Gamma_0=(q/h)^2 a^4 \Omega_\mathrm{P}^2 \Sigma_\mathrm{P},
\end{equation}
which corresponds to the total torque of the unperturbed disc at the initial planet location of $5\,\au$ \citep[see e.g.][]{d'AngeloLubow2010, DuermannKley2015}. It is a function of the planet to stellar mass ratio $q$, the disc aspect ratio $h$, $\Sigma_\mathrm{P}$=$\Sigma_0$ at $5\,\au$, the orbital separation of the planet $a$ and $\Omega_\mathrm{P}=\sqrt{GM_\star/a^3}$.
The torques were smoothed using a third order polynomial Savitzky-Golay filter \citep{SavitzkyGolay1964} with a window length of 11. 
Fluctuations are observed in the torque evolution as the planet moves closer to the central star, in particular for the transition disc cases (right column). The cause of these fluctuations is the interaction between vortices developing at the gap edge and the photoevaporative wind with the planet, which is discussed in more detail in Appendix~\ref{sec:appendix_bumps}.

It is important to note that the transition from a primordial to a transition disc is smooth rather than sharp. While the total torques in the former are the result of the counteracting effect from the torques interior and exterior to the planet, as the disc is slowly developing an inner cavity, the torques in the outer disc will start to dominate over those from interior to the planet. In the extreme limit of a transition disc with an evacuated inner cavity, the surface density will have a sharp cutoff at the planet location as the gap-crossing material will be either quickly accreted by the planet or be photoevaporated by the star. 
At this point, the planet will only be subject to the torques from the outer disc, which we refer to as `one-sided torques'.

It can be inferred from Fig.~\ref{fig:torques} that the absolute value of the Lindblad torques is significantly larger than those of the corotation torques. This is to be expected as the planets in our study are massive enough to open gaps in the surface density, therefore depleting the material in the corotation region. 
This becomes even more prominent in the transition discs since the removal of material due to the direct irradiation from the central star adds up to the planet depleting gas in its immediate surrounding, so that material close to the planet cannot complete a horseshoe orbit. 
Consequently, as photoevaporation depletes the regions close to the planet location, the corotation and Lindblad torques approach zero, and the planet is parked.

\subsection{Impulse approximation versus full 2D treatment}
\label{sec:results_impulseapprox_vs_fargo}

\begin{figure*}
\centering
\includegraphics[width=\hsize]{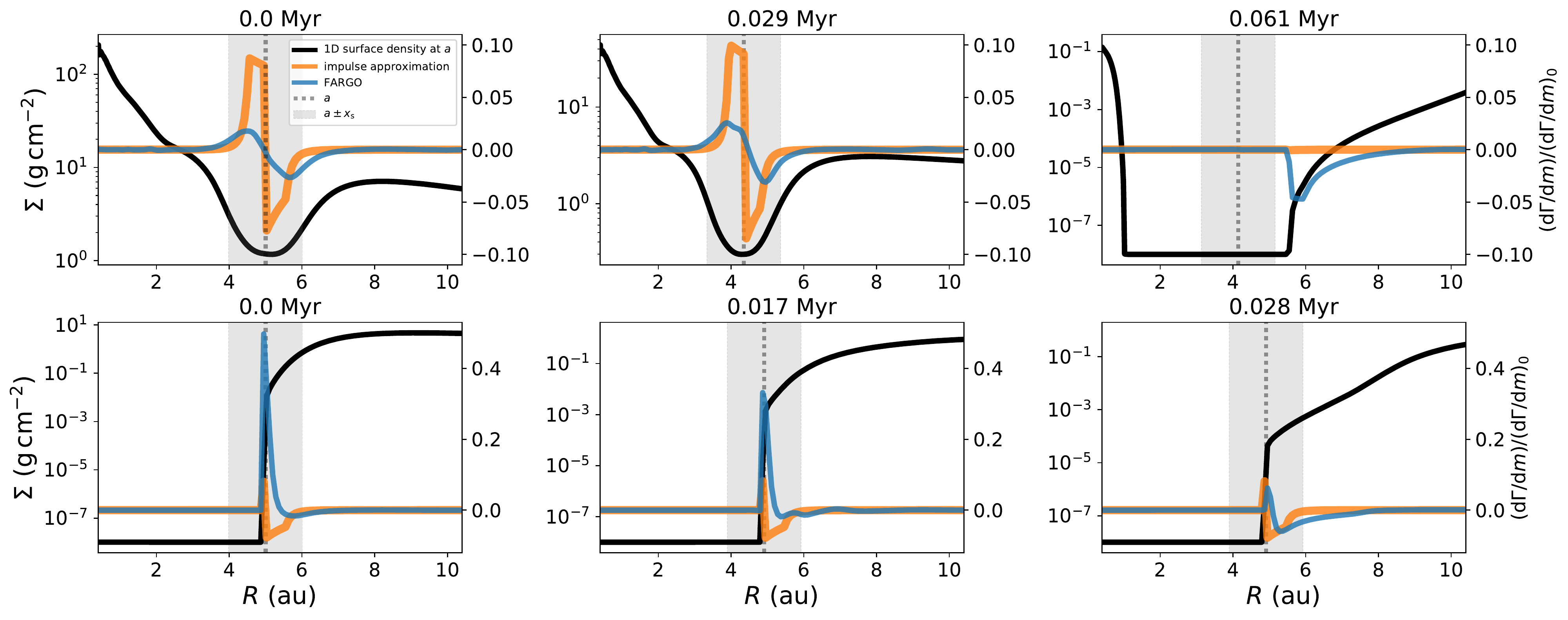}
  \caption{Comparison of the 1D surface density profile (solid black line) of the $\Sigma_0=10\,\gcm$ and $\log_{10}(\Lx/\ergs)=29.54$ primordial (top panels) and transition disc (bottom panels) with the radial distribution of the torques per unit disc mass determined from \fargo\ (blue) and the impulse approximation (orange). The torques were normalised with the normalisation factor $(\mathrm{d}\Gamma/\mathrm{d}m)_0$ (cf. Eq.~\ref{eq:gamma0_mass}). The dotted line shows the planet location at the given timestep, while the shaded area encompasses $a\pm\xs$.}
     \label{fig:torques_radial}
\end{figure*}

Finally, we investigate why the impulse approximation is leading to contradicting results for transition discs compared to migration rates obtained from \fargo. 
The two major differences between \spock\ and \fargo\ are the disc evolution and the treatment of planet migration. Due to the different disc evolution (and surface density profile), it is not useful to directly compare the torques from both approaches. Therefore, in order to isolate the effect of planet migration, we computed the torques from the impulse approximation (cf. Eq.~\ref{eq:impulse_torques}) using the 1D averaged \fargo\ outputs. This way we focus only on the differences in planet migration, which is the aim of this paper.

Fig.~\ref{fig:torques_radial} shows the azimuthally averaged surface density profile of the primordial and transition discs in comparison to the radial distribution of the torques per unit disc mass $\mathrm{d}m$ directly determined from the \fargo\ outputs as well as through the impulse approximation.
The comparison is made at three different stages of disc evolution, namely when the planet is released (first column), in the advanced stage of disc evolution (middle column) as well as right before disc dispersal (right column). 
The radial distribution of the specific torques are plotted in units of the normalisation factor determined by \citet{d'AngeloLubow2010}: 

\begin{equation}
    \label{eq:gamma0_mass}
    \left(\frac{\mathrm{d}\Gamma}{\mathrm{d}m}\right)_0 = q^2 h^{-4} a^2 \Omega_\mathrm{P}^2.
\end{equation}

For the primordial discs (top row), the torques determined from the impulse approximation strongly overestimate the contribution of the corotation torques close to the planet (exerted from the region $a\pm\xs$), while it underestimates the influence of the Lindblad torques at larger radii.
However, as previously shown in Fig.~\ref{fig:torques}, the corotation torques are of minor importance to giant planet migration as the planet will carve a deep gap into the disc, and therefore deplete this region of material.
The absolute values of the torques from \fargo\ are overall smaller than the ones from the impulse approximation.  However, the magnitude of the Lindblad torques, which ultimately drive planet migration, is larger especially close to the planet. The difference between the calculated Lindblad torques in the two approaches is coming mainly from the lack of 2D structures, such as spiral density waves in the averaged density profiles as was previously discussed.  

For the transition discs, the gas cannot perform complete horseshoe orbits since the material is removed as it crosses the planet location. The observed behaviour for the torques differs strongly between the two approaches. There are negative torques just outside the planet location pushing the planet inwards for the impulse approximation, while in \fargo\ strongly positive torques develop at the location of the gap edge, effectively acting as a planet trap that will prevent any further inward migration of the planet \citep[cf.][]{LiuOrmelLin2017}. As the disc evolves and the photoevaporative wind depletes the gap edge this contribution becomes negligible.
Here, it is important to note that our previous statement about the corotation torques not being important to giant planet migration still holds. The reason for this is that the torques shown in Fig.~\ref{fig:torques_radial} correspond to the specific torques, i.e. the torques were renormalised by the local disc mass $\mathrm{d}m=\Sigma  \,\mathrm{d}A = 2\pi r \Sigma \, \mathrm{d}r$. Therefore, while the absolute value of the torques exerted at the gap edge are significantly stronger than in the primordial disc, the local disc mass is much more depleted and the resulting effective torque is weaker as can be inferred from Fig.~\ref{fig:torques}.

\begin{figure*}
\centering
\includegraphics[width=\hsize]{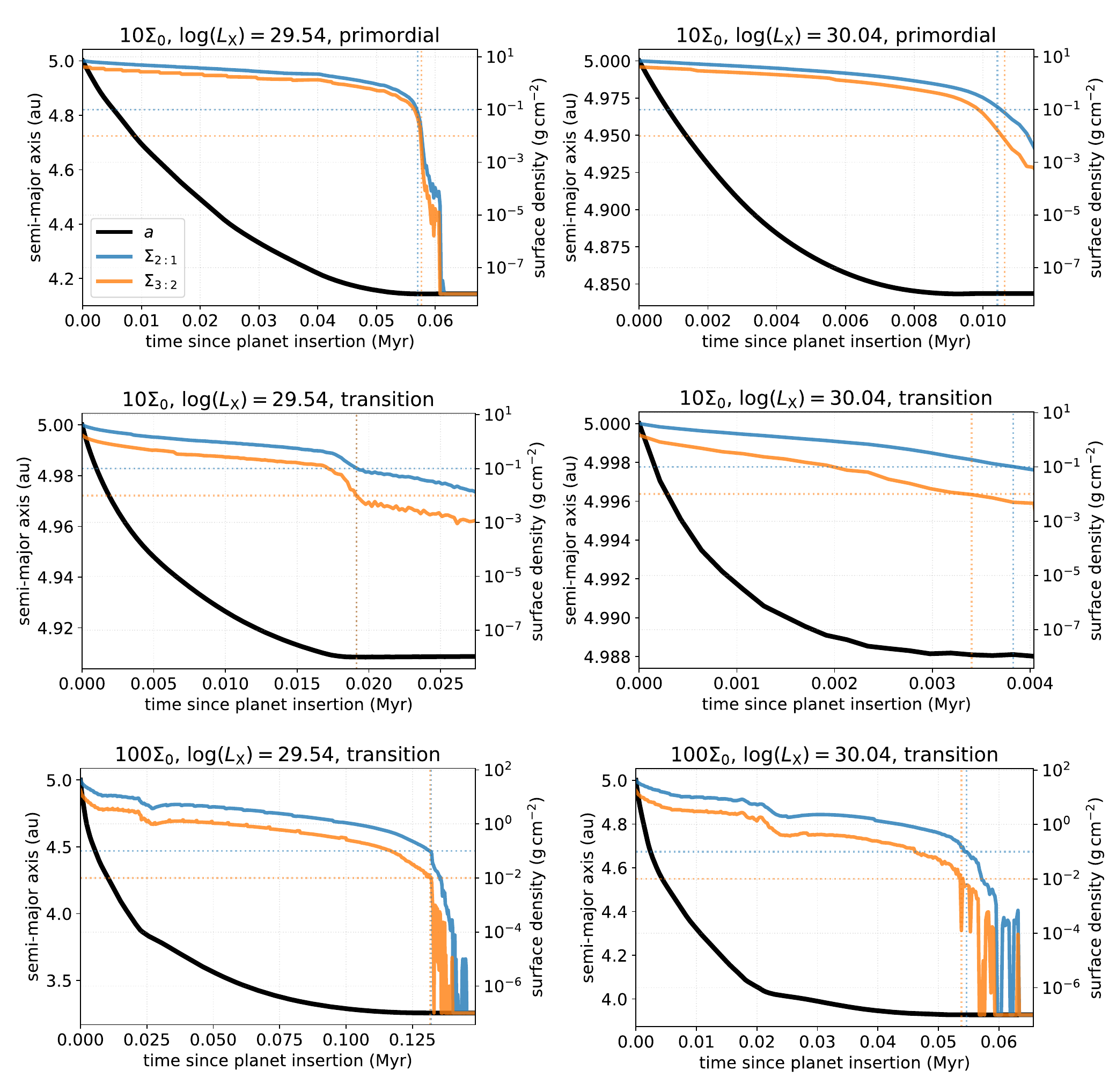}
  \caption{Comparison of the semi-major axis evolution (solid black) in the \fargo\ simulations vs. the azimuthally averaged surface density at the 2:1 (solid blue) and 3:2 (solid orange) mean motion resonance location in the corresponding disc. Only the runs in which the planet was parked due to planet-disc interactions are shown. The horizontal dotted lines are drawn at $\Sigma_\mathrm{2:1} = 10^{-1}\,\gcm$ and $\Sigma_\mathrm{3:2} = 10^{-2}\,\gcm$ and cross the corresponding vertical line where the planet is considered to be parked fully.}
     \label{fig:sigma_resonance}
\end{figure*}

In summary, our 2D \fargo\ simulations show that planets that are embedded in transition discs with an evacuated cavity inside the planetary orbit migrate significantly less than they do in primordial discs. This is in contradiction with the results obtained by the impulse approximation in 1D, which predicts an accelerated inward migration for planets in discs with inner cavities.
This suggests that the classic impulse approximation is not suitable to model the migration of planets in such discs. This has important consequences for 1D planet population synthesis calculations that employ the impulse approximation to calculate migration rates of giant planets.

\subsection{Proposed fix for the impulse approximation in 1D models}
\label{sec:results_fix}

An easy fix to the impulse approximation in 1D models could be to stop giant planet migration, once the disc inside the planetary orbit is depleted of gas. However, as shown in Fig.~\ref{fig:comparison_SPOCK_FARGO}, some modest migration of the planet embedded in a transition disc can be observed, if the outer disc is still rather massive. 
Stopping the planet once the disc inside the planet is depleted would likely result in too large parking radii of the planets.
A better approximation that could be therefore used in such simulations is derived using Fig.~\ref{fig:sigma_resonance}. Here we compare the semi-major axis evolution from the \fargo\ runs with the azimuthally averaged surface density that was determined at the 3:2 and 2:1 mean motion resonance locations. Both are located outside of the planetary orbit and are one of the main contributors to the outer Lindblad torques. The $\Sigma_0=100\,\gcm$ runs of the primordial discs are excluded from this analysis as the planets reached the inner boundary in these simulations and were effectively not parked due to planet-disc interactions. 
We find that once the surface density reaches approximately $0.01\,\gcm$ at the 3:2 mean motion resonance location and around $0.1\,\gcm$ at the 2:1 mean motion resonance, planet migration can be considered to have stopped fully. 
These threshold values are independent of the initial disc mass as well as the applied X-ray luminosity tested in this work and therefore appear to be a robust proxy. 
However, while their eligibility for even higher disc masses still needs to be tested, it is not expected to be of relevance to realistic planet formation models. 
In 1D treatments of migrating planets using the impulse approximation, we therefore suggest to stop giant planet migration once the disc inside the planet is depleted and the surface density at the 3:2 mean motion resonance location (as this is closer to the planet than the 2:1 mean motion resonance) becomes less than $0.01\,\gcm$, ensuring that the outer disc has become depleted enough to not continue pushing the planet inside.

%% file: 04discussion.tex
\section{Numerical limitations}
\label{sec:discussion}

This section will discuss the limitations of the numerical models employed in this study.

\subsection{Viscous boundary conditions}
\label{sec:discussion_boundary}

Due to the reduced complexity of the disc structure in one dimension, it was possible to model a much larger disc extent with \spock\ than in the 2D simulations with \fargo, which only included disc radii between $0.4-25\,\au$. 
In such models, the choice of boundary conditions may have a significant impact on the final outcome of a simulation. While we have tested different configurations of open and closed boundaries, we found that viscous outflow boundary conditions \citep{Kley+2008} provide the most realistic setup as they impose a steady-state accretion flow from the outer to the inner disc as well as from the innermost disc onto the star (cf. Sect.~\ref{sec:methods_fargo}).
The velocity of this flow is set by the viscosity parameter $\alpha$, for which we used the same value of $\alpha=6.9\times10^{-4}$ both in 1D and 2D. However, as could be inferred from Fig.~\ref{fig:sigma_comp}, the discs in our model have reached a stable state before the planets were released, confirming that the choice of viscous boundary conditions is indeed appropriate for our study.

\subsection{Planetary accretion}
\label{sec:discussion_leakage}

In the classical framework of the impulse approximation \citep{LinPapaloizou1986}, no material can cross the planetary gap so that the planet is locked inside the gap, migrating approximately on the viscous timescale. Hydrodynamical simulations have however shown that gaps formed by giant planets are more comparable to `leaky dams' that do allow a certain mass-flow across the planetary orbit \citep{LubowD'Angelo2006, Duffell+2014, Dempsey+2020}. \citet{DuermannKley2015} further showed that, for some scenarios, this mass-flow can be entirely stopped if the planet is additionally accreting the gap-crossing material, therefore fully cutting off the inner from the outer disc. 

In \spock, planetary accretion is modelled following the prescription derived by \citet{VerasArmitage2004}, which was also used in comparable studies \citep{AA09, AP12, ER15}. It is well known that the magnitude of this mass-flow can strongly affect the migration rate of the planet. While the fitting formula that is used in \spock\ is based on state-of-the-art numerical simulations of planet-disc interactions \citep{Lubow+1999, D'Angelo+2002, LubowD'Angelo2006}, its exact form is not known and may therefore carry large uncertainties. 

In contrast, the simulations performed with \fargo\ do not include planetary accretion. 
This may likely be a caveat of our analysis, however, this approach was chosen to isolate the effect of XPE onto giant planet migration.
Furthermore, \citet{Robert+2018} find that the accretion of material onto the planet in discs with low viscosity (as it is the case in our study) should not have any significant impact on type~II migration, suggesting that giant planets embedded in such discs should migrate slowly. 
Studying the mass-flow across the planetary orbit in photoevaporating discs is not within the scope of this paper, however, due to its likely significance to giant planet migration in high-viscosity discs, it should be investigated in more detail in future studies.

%% file: 05conclusion.tex
\section{Conclusions}
\label{sec:conclusion}

In this paper we have tested the impulse approximation as a 1D treatment of giant planet migration in evolving protoplanetary discs subject to photoevaporation. Our conclusions based on our comparison of 1D and 2D simulations can be summarised as follows. 

\begin{enumerate}
    \item The impulse approximation in 1D can roughly reproduce the migration history predicted by more complex 2D simulations for giant planets embedded in primordial discs with or without photoevaporation. Despite some quantitative differences, the effect on large population synthesis models is not expected to be dramatic. One exception to this is the case of planets embedded in high-mass discs with more vigorous winds. Here, the parking effect of photoevaporation is more enhanced in the 1D simulations compared to the 2D simulations. This is due to the strong non-axisymmetric density enhancements, such as spiral waves in the 2D simulations, that are absent in the 1D simulations. These regions are more resilient to photoevaporation and provide an additional contribution to the torques that are driving migration. 
    
    \item 1D simulations show higher migration rates in the case of planets embedded in transition discs, in which the material inside the planetary orbit has been cleared. On the contrary, 2D simulations of equivalent systems show that migration is effectively stopped as soon as the material inside the planetary orbit is cleared. 
    
    \item Planet population synthesis calculations should account for the above in order to accurately reproduce the orbital distribution of planets at the end of the disc clearing phase. A simple approach for 1D codes using the impulse approximation is to halt planet migration as soon as the material inside the planetary orbit has been cleared and the azimuthally averaged surface density becomes less than $0.01\,\gcm$ at the 3:2 mean motion resonance location of the planet. 
\end{enumerate}

A quantitative estimate of the magnitude of the error introduced by 1D migration prescriptions is beyond the scope of this paper but will be attempted in future work.